\documentclass[preprints,article,accept,moreauthors,pdftex]{mdpi} 
\pdfoutput=1
\usepackage{latexsym}
\usepackage{amsmath,amssymb}

\newcommand{\lc}{\varepsilon}

\newcommand{\ds}{\displaystyle}

\newcommand{\D}{\mathrm{d}}
\newcommand{\pb}[2]{\{\,  {#1} \, ,\, {#2} \,\}  }

\newcommand{\killing}[2]{ {\langle #1 \, , #2 \rangle } }
\newcommand{\wkilling}[2]{ {\langle #1 \wedge #2 \rangle } }
\newcommand{\poisson}[2]{ {\{ #1 \, , #2 \} } }

\newcommand{\realni}{\ensuremath{\mathbb{R}}}

\newcommand{\cA}{{\cal A}}

\newcommand{\cF}{{\cal F}}
\newcommand{\cG}{{\cal G}}
\newcommand{\cH}{{\cal H}}

\newcommand{\cM}{{\cal M}}

\newcommand{\cS}{{\cal S}}

\usepackage{environ}
\NewEnviron{myequation}{%
\begin{equation}
\scalebox{0.88}{$\BODY$}
\end{equation}}

\firstpage{1} 
\pubvolume{xx}
\issuenum{1}
\articlenumber{5}
\pubyear{2020}
\copyrightyear{2020}
\history{}

\Title{Hamiltonian Analysis for the Scalar Electrodynamics as $3BF$ Theory}
\Author{Tijana Radenkovi\' c $^{*,\dagger}$\orcidA{} and Marko Vojinovi\' c $^{\dagger}$\orcidB}
\AuthorNames{Tijana Radenkovic and Marko Vojinovic}

\address[1]{
 Institute of Physics, University of Belgrade, Pregrevica 118, 11080 Belgrade, Serbia}

\corres{\hangafter=1 \hangindent=1.05em \hspace{-0.82em} Correspondence: rtijana@ipb.ac.rs}

\firstnote{These authors contributed equally to this work.}

\abstract{The higher category theory can be employed to generalize the $BF$ action to the so-called $3BF$ action, by passing from the notion of a gauge group to the notion of a gauge $3$-group. The theory of scalar electrodynamics coupled to Einstein--Cartan gravity can be formulated as a constrained $3BF$ theory for a specific choice of the gauge $3$-group. The complete Hamiltonian analysis of the $3BF$ action for the choice of a Lie $3$-group corresponding to scalar electrodynamics is performed. This~analysis is the first step towards a canonical quantization of a $3BF$ theory, an important stepping stone for the quantization of the complete scalar electrodynamics coupled to Einstein--Cartan gravity formulated as a $3BF$ action with suitable simplicity constraints. It is shown that the resulting dynamic constraints eliminate all propagating degrees of freedom, i.e., the $3BF$ theory for this choice of a $3$-group is a topological field theory, as expected.}


\begin{document}

\setcounter{section}{0}
\section{Introduction\label{section0}}

The vast majority of physics community agrees that the quantum theory of gravity is necessary, even if they disagree on the quantization approach. The theory of loop quantum gravity is one of the well-formulated possible candidates for the desired theory of quantum gravity \cite{RovelliBook,RovelliVidottoBook,Thiemann2007}. There are two approaches within the theory---the canonical 
and the covariant quantization method. The covariant quantization method is focused on obtaining a generating functional, by considering a triangulated spacetime manifold and defining the functional as a state sum over all configurations of a field living on simplices of the triangulation \cite{RovelliVidottoBook}.

One of the key tools in the covariant quantization approach is the so-called $BF$ theory. Given a Lie group $G$ and its corresponding Lie algebra $\mathfrak{g}$, one considers a $\mathfrak{g}$-valued connection $1$-form $A$, and~its corresponding field strength $2$-form $F \equiv \D A + A\wedge A$. Multiplying $F$ with a $\mathfrak{g}$-valued Lagrange multiplier $2$-form $B$ and integrating over a four-dimensional spacetime manifold $\cM$, one obtains the action of the $BF$ theory,
$$
S_{BF}[A,B] = \int_{\cM} \wkilling{B}{F}_{\mathfrak{g}}\, ,
$$
where $\killing{\_}{\_}_{\mathfrak{g}}$ is a $G$-invariant non-degenerate symmetric bilinear form. The $BF$ theory derives its name from the symbols $B$ and $F$ for the Lagrange multiplier and the field strength present in the action. As it is defined, the $BF$ theory is topological, containing no local propagating degrees of freedom. Therefore, for the purpose of building physically relevant actions, attention usually focuses not on the pure $BF$ theory, but rather on the theory with constraints. The constrained $BF$ models are based on deformations of the $BF$ theory \cite{BFgravity2016}, by adding constraints to the topological $BF$ action that promote some of the gauge degrees of freedom into physical ones. The well known example is the Plebanski model for general relativity \cite{plebanski1977}. Constrained $BF$ models represent a starting point in the spinfoam approach to the construction of quantum gravity models \cite{RovelliVidottoBook}.

The main shortcoming of building a quantum gravity model using a $BF$ theory is the fact that it is very hard, if not impossible, to write the action for matter fields (specifically scalar and fermion fields) in the form of a constrained $BF$ theory. Thus, the spinfoam quantization method is limited to pure gravity, and the problem of consistently coupling matter fields to gravity in this framework becomes highly nontrivial. One of the proposed ways to circumvent this issue is to generalize the notion of a $BF$ theory using the mathematical apparatus of higher category theory.

The higher category theory \cite{BaezHuerta2011} can be employed to generalize the $BF$ action to the so-called $nBF$ action, by passing from the notion of a gauge group to the notion of a gauge $n$-group (for~a comprehensive review of $n$-groups see for example \cite{Miranda2000}, and also Appendix \ref{ApendiksCnovi}). Specifically, the notion of a $3$-group in the framework of higher category theory is introduced as a $3$-category with only one object where all the morphisms, $2$-morphisms and $3$-morphisms are invertible. Based on this generalization, recently a constrained $3BF$ action has been introduced, which describes the full Standard Model coupled to Einstein--Cartan gravity \cite{Radenkovic2019}.

As a first step to the study of the Hamiltonian structure of such theories, in this work, we discuss the simplest nontrivial toy example, namely the theory of scalar electrodynamics coupled to gravity. The standard way to define scalar electrodynamics coupled to gravity is by the action:
\begin{equation} \label{eq:skalarnaelektrodinamika}
S=\int \D^4x \sqrt{-g} \left[ -\frac{1}{16\pi l_p^2} R -\frac{1}{4} g^{\mu\rho} g^{\nu\sigma} F_{\mu\nu}F_{\rho\sigma} + g^{\mu\nu} \nabla_{\mu}\phi^* \nabla_{\nu}\phi - m^2 \phi^*\phi \right]\, .
\end{equation}
Here, $g_{\mu\nu}$ is the spacetime metric, $g\equiv \det (g_{\mu\nu})$ is its determinant, $R$ is the corresponding curvature scalar, and $l_p$ is the Planck length, its square being equal to the Newton's gravitational constant, $l_p^2 = G$, in the natural system of units $\hbar=c=1$. The total covariant derivative $\nabla_{\mu}$ of the complex scalar field $\phi$ is defined as $\nabla_{\mu} \phi = ( \partial_{\mu} + iq A_{\mu} ) \phi$, and thus coupled to the electromagnetic potential $A_{\mu}$ via the coupling constant $q$ (the electric charge of the field $\phi$). See Appendix \ref{ApendiksA} for more detailed notation. In~the next section, we will reformulate this model as a classically equivalent constrained $3BF$ theory for a specific choice of the gauge $3$-group. Moreover, for reasons of simplicity, in the Hamiltonian analysis, we will focus only on the topological sector, disregarding the simplicity constraints. The Hamiltonian structure of the theory is important for various reasons, primarily for the canonical quantization program.

The layout of the paper is as follows. In Section \ref{section1}, we introduce the $3$-group structure corresponding to the theory of scalar electrodynamics coupled to Einstein--Cartan gravity and the corresponding constrained $3BF$ action. Section \ref{section2} contains the Hamiltonian analysis for the topological, $3BF$ sector of the action, with the resulting first-class and second-class constraints present in the theory, and their mutual Poisson brackets. In Section \ref{section3}, we analyze the Bianchi identities that the first-class constraints satisfy, which enforce restrictions in the sense of Hamiltonian analysis, and reduce the number of independent first-class constraints present in the theory. Section \ref{section4} focuses on the counting of the dynamical degrees of freedom present in the theory, based on the results from Sections \ref{section2} and~\ref{section3}. Encouraged by these results, in Section \ref{section5}, we construct the generator of the gauge symmetries for the topological theory and we find the form variations of all variables and their canonical momenta. Finally, Section \ref{section6} is devoted to the discussion of the results and the possible future lines of research. The Appendices contain various technical details.
 
The notation and conventions are as follows. The local Lorentz indices are denoted by the Latin letters $a,b,c,\dots$, take values $0,1,2,3$, and are raised and lowered using the Minkowski metric $\eta_{ab}$ with signature $(-,+,+,+)$. Spacetime indices are denoted by the Greek letters $\mu,\nu,\dots$, and are raised and lowered by the spacetime metric $g_{\mu\nu} = \eta_{ab} e^a{}_{\mu} e^b{}_{\nu}$, where $e^a{}_{\mu}$ are the tetrad fields. The inverse tetrad is denoted as $e^{\mu}{}_a$, so that the standard orthogonality conditions hold: $e^a{}_{\mu} e^{\mu}{}_b = \delta^a_b$ and $e^a{}_{\mu} e^{\nu}{}_a = \delta^{\nu}_{\mu}$. When needed, spacetime indices will be split into time and space indices, denoted with a $0$ and lowcase Latin indices $i,j,\dots$, respectively.  All other indices that appear in the paper are dependent on the context, and their usage is explicitly defined in the text where they appear. The antisymmetrization over two indices is introduced with the factor one half that is $A_{[a_1|a_2\dots a_{n-1}|a_n]}=\frac{1}{2}\left(A_{a_1a_2\dots a_{n-1}a_n}-A_{a_{n}a_2\dots a_{n-1}a_1}\right)$, and the total antisymmetrization is introduced as $A_{[a_1\dots a_n]}=\frac{1}{n!}\sum_{\sigma\in S_n}(-1)^{\mathrm{sign}(\sigma)}A_{a_{\sigma(1)}\dots a_{\sigma(n)}}$.

\section{Scalar Electrodynamics as a Constrained \boldmath$3BF$ Action\label{section1}}

Let us begin by providing a short introduction into the construction and structure of a $3BF$ theory, after which we will impose appropriate simplicity constraints, in order to obtain the equations of motion for scalar electrodynamics coupled to gravity.

As was discussed in detail in \cite{Radenkovic2019}, one formulates a topological $3BF$ action by specifying a particular gauge Lie $3$-group. It has been proved that any strict $3$-group is equivalent to a $2$-crossed \mbox{module \cite{martins2011,Wang2014}}. A gauge theory for the manifold $\mathcal{M}_4$ and $2$-crossed module $(L\stackrel{\delta}{\to} H \stackrel{\partial}{\to}G\, , \rhd\, , \{\_ \, ,\_\})$ can be constructed for the following choice of the three Lie groups as: $$G=SO(3,1)\times U(1)\, ,\quad H=\mathbb{R}^4\, ,\quad L=\mathbb{R}^2\, .$$  The maps $\partial$ and $\delta$ are chosen to be trivial. The action of the algebra $\mathfrak{g}$ on $\mathfrak{h}$ and $\mathfrak{l}$ is chosen as:
\begin{equation} \label{eq:rhd}
\begin{array}{ll}
M_{ab}\rhd P_c=\rhd_{ab,c}{}^d\,P_d=\delta_{[a|}{}^d\eta_{|b]c}\,P_d=\eta_{[b|c}\,P_{|a]}\, , &\quad T\rhd P_a = 0\, ,\\ M_{ab}\rhd P_A = 0\, , &\quad T\rhd P_A= \rhd_A{}^B \,P_B
\end{array}
\end{equation}
where $M_{ab}$ denote the six generators of $\mathfrak{so}(3,1)$, $T$ is the sole generator of $\mathfrak{u}(1)$, $P_a$ are the four generators of $\mathbb{R}^4$ and $P_A$ are the two generators of $\mathbb{R}^2$. In the previous expression, the action of the algebra $\mathfrak{u}(1)$ on the algebra $\mathbb{R}^2$ is defined via $$\rhd_A{}^B=iq\,\begin{bmatrix} 1 & 0 \\ 0 & -1
\end{bmatrix}.$$

The action of the algebra $\mathfrak{g}$ on itself is by definition given via the adjoint representation and, for the choice $\mathfrak{g}=\mathfrak{so}(3,1)\times\mathfrak{u}(1)$, one obtains
\begin{equation}\label{eq:rhdc}
    \begin{array}{c}
      M_{ab}\rhd M_{cd}={\rhd_{ab\, ,cd}}^{ef}\,M_{ef}={f_{ab\, ,cd}}^{ef}\,M_{ef}=\eta_{ad}M_{bc}+\eta_{bc}M_{ad}-\eta_{ac}M_{bd}-\eta_{bd}M_{ac}\, ,\vphantom{\ds\int}\\   M_{ab}\rhd T =0\, , \quad T\rhd M_{ab}=0\, ,\quad T\rhd T=0\, ,\vphantom{\ds\int}\\
    \end{array}
\end{equation}
as the consequence of the direct product structure and the Abelian nature of the subgroup $U(1)$. The~Peiffer lifting
\begin{displaymath}
\{ \_ \, , \_ \} : H\times H \to L
\end{displaymath}
is also trivial, i.e., all the coefficients $X_{ab}{ }^A$ are equal to zero:
\begin{equation} \label{eq:peiffercoef}
\{ P_a\, ,P_b\} \equiv X_{ab}{}^A T_A = 0\, .
\end{equation}
Given Lie algebras $\mathfrak{g}$, $\mathfrak{h}$, and $\mathfrak{l}$, one can introduce a $3$-connection $(\alpha, \beta,\gamma)$ given by the algebra-valued differential forms $\alpha \in \cA^1(\cM_4\, ,\mathfrak{g})$, $\beta \in \cA^2(\cM_4\, ,\mathfrak{h})$ and $\gamma \in \cA^3(\cM_4\, ,\mathfrak{l})$. The corresponding fake $3$-curvature $(\cal F\, , G\, , H)$ is then defined as:
\begin{equation}\label{eq:3krivine}
    \cF = \D \alpha+\alpha \wedge \alpha - \partial \beta \, , \quad \quad \cG = \D \beta + \alpha \wedge^\rhd \beta - \delta \gamma\, , \quad \quad \cH = \D \gamma + \alpha\wedge^\rhd \gamma + \{\beta \wedge \beta\} \, ,
    \end{equation}
see \cite{martins2011, Wang2014} for details.
For this specific choice of a $3$-group, where $\alpha=\omega+A$, given by the algebra-valued differential forms $\omega \in \cA^1(\cM_4\, ,\mathfrak{so}(3,1))$, $A \in \cA^1(\cM_4\, , \mathfrak{u}(1))$, $\beta \in \cA^2(\cM_4\, ,\realni^4)$ and $\gamma \in \cA^3(\cM_4\, ,\realni^2)$, the corresponding $3$-curvature $(\cal F\, , G\, , H)$ is defined as
\begin{equation}
\begin{array}{lclcl}
    \ds \cF&=& R^{ab}M_{ab}+ F T&=& \big(\D \omega^{ab}+\omega^{a}{}_c\wedge \omega^{cb}\big)M_{ab}+\D A\;T\, ,\vphantom{\ds\int}\\
    \ds \cG &=& \cG^a P_a & = & \big(\D \beta^a + \omega^{a}{}_b\wedge \beta^b \big) P_a \, ,\vphantom{\ds\int}\\
    \ds \cH &=& \cH^A P_A &=& \big( \D \gamma^A+ \rhd_B{}^A A\wedge \gamma^B\big) P_A\, .\vphantom{\ds\int}\\
\end{array}
\end{equation}
Note that the connection $\omega^{ab}$ is not present in the last expression, as follows from the definition of the action $\rhd$ and the Peiffer lifting $\{\_\, ,\_\}$, see Equations~(\ref{eq:rhd}) and~(\ref{eq:peiffercoef}):
\begin{equation}
    \begin{array}{lcl}
\mathcal{H}&=&\D \gamma + \alpha\wedge^\rhd \gamma + \{\beta \wedge \beta\}\vphantom{\ds\int}\\
&=&\D \gamma^A P_A + (\omega^{ab}M_{ab} + AT)\wedge^\rhd (\gamma^A P_A)\vphantom{\ds\int}\\&=& \D \gamma^A P_A + \omega^{ab}\wedge \gamma^A M_{ab}\rhd P_A + A\wedge \gamma^A T\rhd P_A\vphantom{\ds\int}\\
&=& \D \gamma^A P_A + A\wedge \gamma^A \rhd_A{}^B P_B \vphantom{\ds\int}\\
&=& (\D \gamma^A+\rhd_B{}^A A\wedge{ \gamma^B}) P_A\, .\vphantom{\ds\int}\\
    \end{array}
\end{equation}
The coefficients of the differential $2$-forms $F$ and $R^{ab}$, $3$-form $\cG$, and $4$-form $\cH$ are:
\begin{equation}
\begin{array}{lcl}
F_{\mu\nu} &=& \partial_\mu A_\nu -\partial_\nu A_\mu\, ,\vphantom{\ds\int}\\
{{R}^{ab}{}_{\mu\nu}}&=&\partial_{\mu} {\omega^{ab}{}_{\nu}}-\partial_{\nu}{\omega^{ab}{}_{\mu}} + \omega^a{}_c{}_{\mu}\omega^{cb}{}_{\nu}-\omega^a{}_c{}_{\nu}\omega^{cb}{}_{\mu}\, ,\vphantom{\ds\int}\\
{{\cal G}^a{}_{\mu \nu\rho}}&=&\partial_{\mu}\beta^a{}_{\nu \rho}+\partial_{\nu}\beta^a{}_{\rho \mu}+\partial_{\rho}{\beta^a{}_{\mu \nu}} +{\omega}^a{}_b{}_{\mu}\,\beta^b{}_{\nu \rho}+{\omega}^{a}{}_b{}_{\nu}\,\beta^b{}_{\rho \mu}+{\omega}^a{}_b{}_{\rho}\,\beta^b{}_{\mu \nu}\, ,\vphantom{\ds\int}\\
{{\mathcal{H}}^A{}_{\mu \nu \rho \sigma}}&=&\partial_{\mu} {\gamma^A{}_{\nu \rho \sigma}}-\partial_{\nu} {\gamma^A{}_{\rho \sigma\mu}}+\partial_{\rho} {\gamma^A{}_{ \sigma\mu\nu}}-\partial_{\sigma} {\gamma^A{}_{\mu\nu \rho}}\vphantom{\ds\int}\\&&+\rhd_B{}^A A{}_{\mu}{\gamma^B{}_{\nu\rho\sigma}}-\rhd_B{}^A A{}_{\nu}{\gamma^B{}_{\rho\sigma\mu}}+\rhd_B{}^A A{}_{\rho}{\gamma^B{}_{\sigma\mu\nu}}-\rhd_B{}^A A{}_{\sigma}{\gamma^B{}_{\mu\nu\rho}}\, . \vphantom{\ds\int}   
\end{array}
\end{equation}

Now, one can define a gauge invariant $3BF$ action as:
\begin{equation}\label{eq:3BF}
     S_{3BF}= \int_{{\cM}_4} \left( \wkilling{B}{\cF}_{\mathfrak{g}} + \wkilling{C}{\cG}_{\mathfrak{h}}+ \wkilling{D}{\cH}_{\mathfrak{l}}\right)\, ,
\end{equation}
where $B \in \cA^2(\cM_4\, ,\mathfrak{so}(3,1))$, $C \in \cA^1(\cM_4\, ,\realni^4)$ and $D \in \cA^0(\cM_4\, ,\realni^2)$ are Lagrange multipliers. The forms $\killing{\_}{\_}_{\mathfrak{g}}$, $\killing{\_}{\_}_{\mathfrak{h}}$ and $\killing{\_}{\_}_{\mathfrak{l}}$ are $G$-invariant bilinear symmetric nondegenerate forms on $\mathfrak{g}$, $\mathfrak{h}$ and $\mathfrak{l}$, respectively, defined as
$$\killing{M_{ab}}{M_{cd}}_{\mathfrak{g}}=g_{ab,\, cd}\, ,\quad \killing{T}{T}_{\mathfrak{g}}=1\, , \quad \killing{M_{ab}}{T}_{\mathfrak{g}}=0\, , \quad \killing{P_a}{P_b}_{\mathfrak{h}}=g_{ab}\, ,\quad \killing{P_A}{P_B}_{\mathfrak{l}}=g_{AB}\, ,$$
where 
$$
g_{ab,\,cd}=\eta_{a[c|}\eta_{b|d]}\, ,\qquad g_{ab}=\begin{bmatrix} 1 & 0\\ 0 & 1\end{bmatrix}, \qquad g_{AB}=\begin{bmatrix} 0 & 1 \\ 1 & 0 \end{bmatrix}.
$$
Identifying the Lagrange multiplier $C^a$ as the tetrad field $e^a$, and the Lagrange multiplier $D^A$ as the doublet of scalar fields $\phi^A$,
$$\phi= \phi^A P_A = \phi P_1 + \phi^* P_2\, ,$$ 
based on their transformation properties as discussed in \cite{Radenkovic2019,MikovicVojinovic2012}, the Lagrangian of the action~(\ref{eq:3BF}) obtains the form:
\begin{equation}\label{eq:LagrangianAction}
    S_{3BF}= \int_{\cM_4} \D^4 x \,\epsilon^{\mu\nu\rho\sigma}\big(\frac{1}{4}\,B^{ab}{}_{\mu\nu}\,R^{cd}{}_{\rho\sigma}\, g_{ab,\,cd}+\frac{1}{4}\,B_{\mu\nu}F_{\rho\sigma}+\frac{1}{3!}\, e^a{}_\mu\,
\cG^b{}_{\nu\rho\sigma}\, g_{ab}+ \frac{1}{4!} \, \phi^A \cH^B{}_{\mu\nu\rho\sigma} \, g_{AB}\big)\, .
\end{equation}
Varying the action with respect to all the variables, one obtains the equations of motion:
\begin{equation}
\begin{array}{|c|c|} \hline
\text{varied variable} & \text{equation of motion}  \\ \hline\hline
\delta B^{ab}& R_{ab}=0\vphantom{\ds\int} \\ \hline
\delta \omega^{ab} & \nabla B_{ab}-e_{[a|}\wedge \beta_{|b]}=0\vphantom{\ds\int} \\ \hline
\delta e^a & \cG_a=0\vphantom{\ds\int} \\ \hline
\delta\phi^A & \nabla \gamma_A=0\vphantom{\ds\int}  \\ \hline
\end{array}
\qquad
\begin{array}{|c|c|} \hline
\text{varied variable} & \text{equation of motion} \\ \hline\hline
 \delta B & F=0\vphantom{\ds\int} \\ \hline
\delta A & \D B+\phi_A\,\rhd_B{}^A \,\gamma^B=0\vphantom{\ds\int} \\ \hline
\delta \beta^a & \nabla e_a=0\vphantom{\ds\int} \\ \hline
\delta\gamma^A & \nabla \phi_A=0\vphantom{\ds\int} \\ \hline
\end{array}
\end{equation}

Since one is interested in the doublet of scalar fields $\phi^A$ of mass $m$ and charge $q$ minimally coupled to gravity and electromagnetic field, we impose additional simplicity constraint terms to the topological action~(\ref{eq:3BF}), in order to obtain the appropriate equations of motion equivalent to the equations of motion for the action~(\ref{eq:skalarnaelektrodinamika}):
\begin{myequation}\label{eq:scalarelectrodynamics}
\begin{aligned}
 S =\int_{\cM_4} & B^{ab}\wedge R_{ab} + B \wedge F + e_a\wedge \nabla \beta^a + \phi_A \, \nabla \gamma^A \vphantom{\ds\int} \\
 &- \lambda_{ab} \wedge \Big(B^{ab}-\frac{1}{16\pi l_p^2}\varepsilon^{abcd} e_c \wedge e_d\Big)\vphantom{\ds\int} \\
 &+ {\lambda}^A\wedge \Big(\gamma_A - \frac{1}{2} H_{abcA} e^a \wedge e^b \wedge e^c\Big) +\Lambda^{abA}\wedge \Big( H_{abcA}\varepsilon^{cdef}e_d\wedge e_e \wedge e_f- \nabla \phi_A \wedge e_a \wedge e_b\Big) \vphantom{\ds\int} \\
 &+ \lambda\wedge \Big(B-\frac{12}{q}{M_{ab}}e^a\wedge e^b\Big) + {\zeta^{ab}}\Big( {M_{ab}}\varepsilon_{cdef}e^c\wedge e^d \wedge e^e \wedge e^f- F \wedge e_a \wedge e_b \Big)\vphantom{\ds\int} 
 \\
 &-\frac{1}{2\cdot 4!} m^2\phi_A\,\phi^A \varepsilon_{abcd}e^a\wedge e^b \wedge e^c \wedge e^d\vphantom{\ds\int}\, .
\end{aligned}
\end{myequation}
For the notation used here and the equations of motion obtained by varying the action~(\ref{eq:scalarelectrodynamics}), see Appendix \ref{ApendiksA}.

The dynamical degrees of freedom are the tetrad fields $e^a$, the scalar doublet $\phi^A$, and the electromagnetic potential $A$, while the remaining variables are algebraically determined in terms of them, as shown in Appendix \ref{ApendiksA}. The equation of motion for the field $\phi^A$ reduces to the covariant Klein-Gordon equation for the scalar field,
\begin{equation}
\left(\nabla_\mu\nabla^\mu -m^2\right)\phi_A=0\, .
\end{equation}
The differential equation of motion for the field $A$ is:
\begin{equation}
    \nabla_\mu F^{\mu\nu}=j^\nu\, , \quad \quad j^\mu\equiv \frac{1}{2}\Big(\nabla^\nu \phi^A \rhd^B{}_A \phi_B - \phi_A \rhd_B{}^A \nabla^\nu \phi^B\Big)=iq\,\Big(\nabla \phi^*\,\phi-\phi^* \nabla \phi\Big)\, .
\end{equation}
Finally, the equation of motion for $e^a$ becomes:
\begin{equation}
\begin{array}{c}
     \ds R^{\mu\nu}-\frac{1}{2}g^{\mu\nu} R=8\pi l_p^2 \; T^{\mu\nu}\, ,\vphantom{\ds\int} \\
     \ds T^{\mu\nu}\equiv\nabla^\mu \phi_A \, \nabla^\nu \phi^A -\frac{1}{2}g^{\mu\nu} \left(\nabla_\rho \phi_A\, \nabla^\rho \phi^A+m^2\phi_A\,\phi^A \right)-\frac{1}{4q}\left(F_{\rho \sigma}F^{\rho \sigma}g^{\mu \nu}+4F^{\mu \rho}{F_\rho}^{\nu} \right)\, .\vphantom{\ds\int}
\end{array}
\end{equation}

\section{The Hamiltonian Analysis\label{section2}}

The Hamiltonian analysis of the constrained $3BF$ action~(\ref{eq:scalarelectrodynamics}) for scalar electrodynamics is exceedingly complicated to study. A testament to this is the level of complexity of the constrained $2BF$ formulation of general relativity \cite{MOV2019}, which is merely one sector in the action~(\ref{eq:scalarelectrodynamics}). Therefore, in this paper, we will limit ourselves to the topological sector of the theory, namely the unconstrained $3BF$ theory~(\ref{eq:3BF}), which consists of the terms in the first row of Equation~(\ref{eq:scalarelectrodynamics}), and is written in full detail in Equation~(\ref{eq:LagrangianAction}). One~should be aware that this restriction changes various properties of the theory. Namely, the simplicity constraints (everything but the first row in Equation~(\ref{eq:scalarelectrodynamics})) substantially modify the dynamics of the theory---they increase the number of local propagating degrees of freedom of the theory, a property that was known since the original Plebanski model \cite{plebanski1977}. On the other hand, the unconstrained $3BF$ theory~(\ref{eq:3BF}) is important even in its own right, and the Hamiltonian analysis may give important insight into the structure of both the unconstrained and the constrained theory.

In what follows, the complete Hamiltonian analysis for the action~(\ref{eq:3BF}) is presented, see \cite{blagojevic2002gravitation} for an overview and a comprehensive introduction of the Hamiltonian analysis. The Hamiltonian analysis for a $2BF$ action is performed in \cite{MikovicOliveira2014, MOV2016, MOV2019, Mikovic2015}.

Under the standard assumption that the spacetime manifold is globally hyperbolic, $\cM_4 = \realni \times\Sigma_3$, the Lagrangian of the action~(\ref{eq:3BF}) has the form:
\begin{equation}\label{eq:Lagrangian}
    L_{3BF}= \int_{{\Sigma}_3} \D^3 \vec{x} \,\epsilon^{\mu\nu\rho\sigma}\big(\frac{1}{4}\,B^{ab}{}_{\mu\nu}\,R^{cd}{}_{\rho\sigma}\, g_{ab,\,cd}+\frac{1}{4}\,B_{\mu\nu}F_{\rho\sigma}+\frac{1}{3!}\, e^a{}_\mu\,
\cG^b{}_{\nu\rho\sigma}\, g_{ab}+ \frac{1}{4!} \, \phi^A \cH^B{}_{\mu\nu\rho\sigma} \, g_{AB}\big)\, .
\end{equation}
The canonical momentum $\pi(q)$ corresponding for the canonical coordinate $q$ from the set of all variables in the theory, $q \in \{ B^{ab}{}_{\mu\nu} , \omega^{ab}{}_\mu , B_{\mu\nu} , A_{\mu} , e^a{}_\mu , \beta^a{}_{\mu\nu} , \phi^A , \gamma^A{}_{\mu\nu\rho} \}$, is obtained as a derivative of the Lagrangian with respect to the appropriate velocity,
$$
\pi(q) \equiv \frac{\delta L}{\delta \partial_0 q}\, ,
$$
giving:
\begin{equation}
\begin{array}{lclclcl}
\ds \pi(B){_{ab}{}^{\mu\nu}} & = &  0\, , \vphantom{\ds\int} & & \ds \pi(\omega){_{ab}{}^{\mu}} & = & \epsilon^{0\mu\nu\rho} B_{ab\nu\rho}\, ,\vphantom{\ds\int}\\
\ds \pi(B){{}^{\mu\nu}} & = & 0\, , \vphantom{\ds\int} & & \ds \pi(A){{}^{\mu}} & = & \ds \frac{1}{2} \epsilon^{0\mu\nu\rho} B_{\nu\rho}\, ,\vphantom{\ds\int}\\
\ds \pi(e){_a{}^\mu} & = & 0 \, ,\vphantom{\ds\int} & & \ds \pi(\beta){_a{}^{\mu\nu}} & = & - \epsilon^{0\mu\nu\rho} e_{a\rho}\, ,\vphantom{\ds\int}\\
\pi(\phi){_A} & = & 0\, ,\vphantom{\ds\int} & & \ds \pi(\gamma){_A{}^{\mu\nu\rho}} & = & \epsilon^{0\mu\nu\rho} \phi_A\, .\vphantom{\ds\int}\\
\end{array}
\end{equation}
Since these momenta cannot be inverted for the time derivatives of the variables, they all give rise to primary constraints:
\begin{equation}
\begin{array}{lclclcl}
P(B){_{ab}{}^{\mu\nu} }& \equiv & \pi(B){_{ab}{}^{\mu\nu}} \approx 0\, , \vphantom{\ds\int} & & P(\omega){_{ab}{}^{\mu}} & \equiv & \pi(\omega){_{ab}{}^{\mu}} - \epsilon^{0\mu\nu\rho} B_{ab \nu\rho} \approx 0\, ,\vphantom{\ds\int} \\
P(B){{}^{\mu\nu}} & \equiv &  \pi(B){{}^{\mu\nu}}\approx 0\, , \vphantom{\ds\int} & & P(A){{}^{\mu}} & \equiv &  \pi(A){{}^{\mu}} -\frac{1}{2} \epsilon^{0\mu\nu\rho} B_{\nu\rho}\approx 0\, ,\vphantom{\ds\int}\\
P(e){_a{}^{\mu}} & \equiv & \pi(e){_a{}^{\mu}}  \approx 0\, , \vphantom{\ds\int} & & P(\beta){_a{}^{\mu\nu}} & \equiv & \pi(\beta){_a{}^{\mu\nu}} + \epsilon^{0\mu\nu\rho}e_{a\rho} \approx 0\, , \vphantom{\ds\int}\\
P(\phi){_{A}{}}& \equiv & \pi(\phi){_{A}{}} \approx 0\, , \vphantom{\ds\int} & & P(\gamma){_A{}^{\mu\nu\rho}} & \equiv & \pi(\gamma)_A{}^{\mu\nu\rho} - \epsilon^{0\mu\nu\rho} \phi_A \approx 0\, . \vphantom{\ds\int}\\
\end{array}
\end{equation}
Here, the symbol ``$\approx$'' denotes the so-called ``weak'' equality, i.e., the equality that holds on a subspace of the phase space determined by the constraints, while the equality that holds for any point of the phase space is referred to as the ``strong'' equality and it is denoted by the symbol ``$=$''. The expressions ``on-shell'' and ``off-shell'' are used for weak and strong equalities, respectively, and henceforth will be used in this paper.

The fundamental Poisson brackets are defined as:
\begin{equation}
\begin{array}{lcl}
\ds \pb{B{^{ab}{}_{\mu\nu}}(x)}{\pi(B){_{cd}{}^{\rho\sigma}(y)}} & = &  4\delta{}^a{}_{[c}\delta{}^b{}_{d]}  \delta{}^\rho{}_{[ \mu} \delta{}^{\sigma}{}_{\nu]} \,\delta^{(3)}(\vec{x}-\vec{y})\, , \vphantom{\ds\int}\\
\ds \pb{\omega{^{ab}{}_{\mu}}(x)}{\pi(\omega){_{cd}{}^{\nu}(y)}} & = &  2\delta{}^a{}_{[c}\delta{}^b{}_{d]}  \delta^\nu{}_{ \mu} \,\delta^{(3)}(\vec{x}-\vec{y})\, , \vphantom{\ds\int}\\
\ds \pb{B{{}_{\mu\nu}}(x)}{\pi(B){{}^{\rho\sigma}(y)}} & = & 2\delta{}^\rho{}_{[ \mu} \delta{}^{\sigma}{}_{\nu]} \,\delta^{(3)}(\vec{x}-\vec{y})\, , \vphantom{\ds\int}\\
\ds \pb{A{{}_{\mu}}(x)}{\pi(A){{}^{\nu}(y)}} & = &   \delta{}^\nu{}_{ \mu} \,\delta^{(3)}(\vec{x}-\vec{y})\, , \vphantom{\ds\int}\\
\ds \pb{e{^a{}_{\mu}}(x)}{\pi(e){_b{}^{\nu}}(y)} & = & \delta{}^a{}_b \delta{}^{\nu}{}_{\mu} \,\delta^{(3)}(\vec{x}-\vec{y})\, , \vphantom{\ds\int}\\
\pb{\beta{^a{}_{\mu\nu}}(x)}{\pi(\beta){_b{}^{\rho\sigma}}(y)} & = & 2\delta{}^a{}_b \,\delta{}^{\rho}{}_{[\mu} \delta{}^{\sigma}{}_{\nu]} \,\delta^{(3)}(\vec{x}-\vec{y})\, , \vphantom{\ds\int}\\
\pb{\phi{^A{}}(x)}{\pi(\phi){_B{}}(y)} & = & \delta{}^A{}_B \,\delta^{(3)}(\vec{x}-\vec{y})\, , \vphantom{\ds\int}\\
\pb{\gamma{^A{}_{\mu\nu\rho}}(x)}{\pi(\gamma){_B{}^{\alpha\beta\gamma}}(y)} & = & 3!\delta{}^A{}_B \,\delta{}^{\alpha}{}_{[\mu} \delta{}^{\beta}{}_{\nu} \delta{}^{\gamma}{}_{\rho]}  \,\delta^{(3)}(\vec{x}-\vec{y})\, .\vphantom{\ds\int}
\end{array}
\end{equation}
Using these relations, one can calculate the algebra between the primary constraints,
\begin{equation}\label{AlgebraPrimarnihVeza}
\begin{array}{lcl}
\ds \pb{P(B){^{ab}{}^{jk}}(x)}{P(\omega){_{cd}{}^i}(y)} & = & 4\epsilon^{0ijk}\,\delta{}^a{}_{[c}\delta{}^b{}_{d]} \,\delta^{(3)}(\vec{x}-\vec{y})\, , \vphantom{\ds\int}\\
\ds \pb{P(B){{}^{jk}}(x)}{P(A){{}^i}(y)} & = & \epsilon^{0ijk} \,\delta^{(3)}(\vec{x}-\vec{y})\, , \vphantom{\ds\int}\\
\ds \pb{P(e){^a{}^{k}}}{P(\beta){_b{}^{ij}}(y)} & = & - \epsilon^{0ijk}\, \delta{}^a{}_b(x)\, \delta^{(3)}(\vec{x}-\vec{y})\, , \vphantom{\ds\int}\\
\ds \pb{P(\phi){^A(x)}}{P(\gamma){_B{}^{ijk}}(y)} & = & \epsilon^{0ijk}\, \delta{}^A{}_B\, \delta^{(3)}(\vec{x}-\vec{y})\, , \vphantom{\ds\int}\\
\end{array}
\end{equation}
while all other Poisson brackets vanish. The canonical on-shell Hamiltonian is defined by
\begin{equation}\label{Hcan}
\begin{aligned}
H_c = \int_{{\Sigma}_3} & \D^3\vec{x} \bigg[\frac{1}{4} \pi(B){_{ab}{}^{\mu\nu}}\,\partial_0 B{^{ab}{}_{\mu\nu}} + \frac{1}{2}\pi(\omega){_{ab}{}^{\mu}} \,\partial_0 \omega{^{ab}{}_{\mu}} + \frac{1}{2} \pi(B){{}^{\mu\nu}}\,\partial_0 B{{}_{\mu\nu}} + \pi(A){{}^{\mu}} \,\partial_0 A{{}_{\mu}}\vphantom{\ds\int} \\&+\pi(e){_a{}^{\mu}} \, \partial_0 e{^a{}_{\mu}}+ \frac{1}{2} \pi(\beta){_a{}^{\mu\nu}}\, \partial_0 \beta{^a{}_{\mu\nu}} + \pi(\phi){_{A}} \, \partial_0 D{^{A}} +\frac{1}{3!} \pi(\gamma){_A{}^{\mu\nu\rho}} \, \partial_0 \gamma{^A{}_{\mu\nu\rho}}\bigg]-L\, .\vphantom{\ds\int}
\end{aligned}
\end{equation}
Rewriting the Hamiltonian~(\ref{Hcan}) such that all the velocities are multiplied by the first class constraints and therefore in an on-shell quantity they drop out, one obtains:
\begin{equation}
\begin{aligned}
H_c = & - \int_{\Sigma_3} \D^3\vec{x}\, \epsilon^{0ijk} \bigg[ \frac{1}{2} B_{ab 0i}\, R{^{ab}{}_{jk}} + \frac{1}{2}B_{0i} F_{jk}+ \frac{1}{6} e_{a 0} \,\cG{^a{}_{ijk}} + \beta{^a{}_{0i}} \nabla_j e_{a k}\vphantom{\ds\int}\\ & + \frac{1}{2} \omega{^{ab}{}_0}\bigg( \nabla_i B{_{ab\,jk}} - e{_{[a|i}}\, \beta{_{|b]jk}}\bigg)+ \frac{1}{2} A_0 \bigg( \partial_i B_{jk} + \frac{1}{3} \,\phi_A \,\rhd_B{}^A\,\gamma^B{}_{ijk}\bigg) + \frac{1}{2}\gamma{^A{_{0ij}}}\nabla_k \phi_A \bigg] \, . \vphantom{\ds\int}
\end{aligned}
\end{equation}
This expression does not depend on any of the canonical momenta and it contains only the fields and their spatial derivatives. By adding a Lagrange multiplier $\lambda$ for each of the primary constraints we can build the off-shell Hamiltonian, which is given by:
\begin{myequation} \label{TotalniHamiltonijan}
\begin{array}{ccl}
H_T & = & \ds H_c \!+\! \int_{\Sigma_3} \D^3\vec{x} \bigg[ \frac{1}{4} \lambda(B){^{ab}{}_{\mu\nu}} P(B){_{ab}{}^{\mu\nu}} + \frac{1}{2} \lambda(\omega){^{ab}{}_{\mu}} P(\omega){_{ab}{}^{\mu}} +\frac{1}{2} \lambda(B){{}_{\mu\nu}} P(B){{}^{\mu\nu}} +  \lambda(A){{}_{\mu}} P(A){{}^{\mu}}  \vphantom{\ds\int}\\
 & & \ds + \lambda(e){^a{}_{\mu}} P(e){_a{}^{\mu}}+ \frac{1}{2} \lambda(\beta){^a{}_{\mu\nu}} P(\beta){_a{}^{\mu\nu}}+ \lambda(\phi){^A} P(\phi){_A}+\frac{1}{3!}\lambda(\gamma){^A{}_{\mu\nu\rho}} P(\gamma){_A{}^{\mu\nu\rho}} \bigg] \, . \vphantom{\ds\int}
\end{array}
\end{myequation}

Since the primary constraints must be preserved in time, one must impose the following~requirement:
\begin{equation}
\dot{P} \equiv \pb{P}{H_T} \approx 0\, ,\label{pcc}
\end{equation}
for each primary constraint $P$. 
By using the consistency condition~(\ref{pcc}) for the primary constraints ${P}(B)_{ab}{}^{0i}$, ${P}(\omega){_{ab}{}^0}$, ${P}(B){}^{0i}$, ${P}(A){{}^0}$, ${P}(e){_a{}^0}$, ${P}(\beta){_a{}^{0i}}$, and ${P}(\gamma){_{A}{}^{0ij}}$,
\begin{equation}
    \begin{array}{cccc}
       \ds \dot{{P}}(B)_{ab}{}^{0i} \approx 0\, , & \ds \dot{{P}}(\omega){_{ab}{}^0}\approx 0\, , &  \ds \dot{{P}}(B){}^{0i} \approx 0\, , & \ds \dot{{P}}(A){{}^0}\approx 0\, ,\\  \dot{{P}}(e){_a{}^0}\,\approx 0\, ,&   \dot{{P}}(\beta){_a{}^{0i}} \approx 0\, , & \dot{{P}}(\gamma){_{A}{}^{0ij}} \approx 0\, , & \vphantom{\ds\int}\\
    \end{array}
\end{equation}
one obtains the secondary constraints $\cS$,
\begin{equation} \label{SekundarneVeze}
\begin{array}{lclclcl}
\cS(R){_{ab}}{}^{i} & \equiv & \epsilon^{0ijk} R{_{ab}{}_{jk}} \approx 0\, , \vphantom{\ds\int} & & \cS(\nabla B)_{ab} & \equiv & \epsilon^{0ijk} \big( \nabla_{i} B{_{ab\,jk}} - e{_{[a|i}}\, \beta{_{|b]}{}_{jk}} \big) \approx 0 \, ,\vphantom{\ds\int} \\
\cS(F){{}^{i}} & \equiv & \frac{1}{2}\epsilon^{0ijk} F{{}_{jk}} \approx 0\, , \vphantom{\ds\int} & & \cS(\nabla B) & \equiv & \frac{1}{2}\epsilon^{0ijk} \big(\partial_i B_{jk} + \frac{1}{3} \,\phi_A \,\rhd_B{}^A\,\gamma^B{}_{ijk}\big) \approx 0\, ,\vphantom{\ds\int} \\
\cS(\cG)_a & \equiv & \frac{1}{6}\epsilon^{0ijk}\cG{_a{}_{ijk}} \approx 0\, ,\vphantom{\ds\int} & & \cS(\nabla e){_a{}^{i}} & \equiv & \epsilon^{0ijk}\nabla_{j} e_{a k} \approx 0\, ,\vphantom{\ds\int} \\
\cS(\nabla \phi){_A{}^{ij}} & \equiv & \epsilon^{0ijk} \nabla_k \phi_A \approx 0\, ,\vphantom{\ds\int} & & & & \\
\end{array}
\end{equation}
while in the case of ${P}(B){_{ab}{}^{jk}}$, ${P}(\omega){_{ab}{}^k}$, ${P}(B){{}^{jk}}$, ${P}(A){{}^k}$, 
${P}(e){_a{}^k}$, ${P}(\beta){_a{}^{jk}}$, ${P}(\phi){_{A}}$ and ${P}(\gamma){_A{}^{ijk}}$ 
the consistency conditions
\begin{equation}
    \begin{array}{cccc}
       \ds \dot{P}(B){_{ab}{}^{jk}} \approx 0\, , & \dot{P}(\omega){_{ab}{}^k}\approx 0\, , & \ds \dot{P}(B){{}^{jk}} \approx 0\, , & \dot{P}(A){{}^k}\approx 0\, , \vphantom{\ds\int}\\ \dot{P}(e){_a{}^k}\,\approx 0\, ,&   \dot{P}(\beta){_a{}^{jk}} \approx 0\, , & \dot{P}(\phi){_{A}} \approx 0\, , & \dot{P}(\gamma){_A{}^{ijk}} \approx 0\, , \vphantom{\ds\int}
    \end{array}
\end{equation}
determine the following Lagrange multipliers:
{
\begin{equation}
\begin{array}{lclclcl}
\lambda(\omega){_{ab}{}^{i}} & \approx & \nabla^i\, \omega{_{ab\,0}}\, , \vphantom{\ds\int} & \ \qquad\  & \lambda(B){}^{ij}&\approx&2\partial^{[i|}\, B{}^{0|j]}+\gamma_A{}^{0ij}\rhd_B{}^A\,\phi^B  \, ,\vphantom{\ds\int}\\
\lambda (A){{}^{i}} & \approx & \partial^i \,A{{}_0}\, , \vphantom{\ds\int} & & \lambda(\beta){_a{}^{ij}} & \approx &  2\nabla^{[i|}\, \beta{_a{}^{0|j]}} - \omega{_{ab}{}^0}\,\beta{^{b\,ij}}\, , \vphantom{\ds\int}\\
\lambda(\phi)^A & \approx &  A^0\,\rhd{}^A{}_B\,\phi^B\, ,\vphantom{\ds\int} & & \lambda(e)_a{}^i & \approx & \nabla^i\, e_a{}^0-\omega{_a{}^{b\,0}}\, e{_b{}^i} \, ,\vphantom{\ds\int} \\
\lambda(B){_{ab}{}^{ij}}&\approx& \multicolumn{5}{l}{ 2\nabla^{[i|} B_{ab}{}^{0|j]}+e_{[a|\,0} \beta_{|b]}{}^{ij} -2 e_{[a|}{}^{[i|} \beta_{|b]}{}^{0|j]}+2\omega_{[a|}{}^c B_{|b]}{}^{c\,ij} \, ,\vphantom{\ds\int} } \\
\lambda(\gamma){}_A{}^{ijk} & \approx & \multicolumn{5}{l}{ -A^0\,\rhd{}_A{}^B\, \gamma_B{}^{ijk} + \nabla^i\gamma_A{}^{0jk} -\nabla^j\gamma_A{}^{0ik}+\nabla^k\gamma_A{}^{0ij}\, .\vphantom{\ds\int} } \\
\end{array}
\end{equation}}
Note that the consistency conditions leave the Lagrange multipliers
\begin{equation}
\lambda(B){^{ab}{}_{0i}}\, , \qquad
\lambda(\omega){^{ab}{}_0}\, , \qquad
\lambda(B){{}_{0i}}\, , \qquad
\lambda(A){{}_0}\, , \qquad
\lambda(e){^a{}_0}\, , \qquad 
\lambda(\beta){^a{}_{0i}}\, , \qquad
\lambda(\gamma){^A{}_{0ij}}
\end{equation}
undetermined. The consistency conditions of the secondary constraints do not produce new constraints, since one can show that
\begin{equation}
    \begin{array}{lclcl}
    \dot{\cS}(R)^{ab}{}^i&=&\{\cS(R)^{ab}{}^i\, , \,H_T\}&=&\omega^{[a|}{}_c{}_0\,\cS(R)^{c|b]i}\, ,\vphantom{\ds\int}\\
    \dot{\cS}(\nabla B)&=&\{\cS(\nabla B), \,H_T\}&=& -\rhd_B{}^A\,\gamma^B{}_{0ij}\,\cS(\nabla \phi)_A{}^{ij}\, ,\vphantom{\ds\int}\\
    \dot{\cS}(\cG)^a&=&\{\cS(\cG)^a\, ,\,H_T\}&=&\beta_b{}_{0k}\, \cS(R)^{ab}{}^k-\omega^{ab}{}_0\, \cS(\cG)_b\, ,\vphantom{\ds\int}\\
    \dot{\cS}(\nabla e)_a{}^i&=&\{\cS(\nabla e)_a{}^i\, ,\,H_T\}&=&e^b{}_0\,\cS(R)_{ab}{}^i-\omega_a{}^b{}_0\, \cS(\nabla e)_b{}^i\, ,\vphantom{\ds\int}\\
    \dot{\cS}(\nabla \phi){}_A{}^{ij}&=&\{\cS(\nabla \phi){}_A{}^{ij}\, ,\,H_T\}&=&\,A_0 \,\rhd{}_A{}^B \cS(\nabla \phi){}_B{}^{ij}\, ,\vphantom{\ds\int}\\
    \dot{\cS}(F){}^i&=&\{\cS(F){}^i\, , \,H_T\}&=&0\, ,\vphantom{\ds\int}\\
    \dot{\cS}(\nabla B)_{ab}&=&\{\cS(\nabla B)_{ab}\, , \,H_T\}&=& \cS(R)_{[a|c}{}^k\, B^c{}_{|b]0k}+ \omega{}_{[a|}{}^c{}_0\cS(\nabla B)_{|b]c}\vphantom{\ds\int}\\&&&&- \beta_{[a|0k}\,\cS(\nabla e)_{|b]}{}^k\vphantom{\ds\int}+e_{[a|0}\,\cS(\cG){}_{|b]}{}\, .\\
    \end{array}
\end{equation}
Then, the total Hamiltonian can be written as
\begin{equation}
\begin{array}{lcll}
  H_T & = & \ds \int_{\Sigma_3} \D^3{\vec{x}} & \ds \bigg[ \frac{1}{2}\lambda(B){_{ab}{}^{0i}} \,\Phi(B){^{ab}{}_i} + \frac{1}{2}\lambda(\omega)_{ab}{}^0 \,\Phi(\omega)^{ab} + \lambda(B){{}^{0i}} \,\Phi(B){{}_i} + \lambda(A)^0 \,\Phi(A) \\
    & & & \ds \phantom{\ds\int} + \lambda(e){_a{}^0} \,\Phi(e)^a + \lambda(\beta){_a{}^{0i}} \,\Phi(\beta){^a{}_{i}}+\frac{1}{2} \lambda(\gamma){_A{}^{0ij}}\Phi(\gamma)^A{}_{ij} \\
    & & & \ds \phantom{\ds\int} - \frac{1}{2}B_{ab 0i} \,\Phi(R)^{abi} - \frac{1}{2}\omega_{ab 0} \,\Phi(\nabla B)^{ab } -B_{ 0i} \,\Phi(F)^{i} - A_{ 0} \,\Phi(\nabla B)\\
    & & & \ds \phantom{\ds\int} - e_{a0}\,\Phi(\cG)^a -\beta_{a0i} \,\Phi(\nabla e)^{a i} - \frac{1}{2}\gamma_{A 0ij}\, \Phi(\nabla \phi)^{Aij} \bigg]\, ,\\
\end{array}
\end{equation}
where  
\begin{equation}
\begin{array}{lclclcl}
\Phi(B){^{ab}{}_i} & = &P(B)^{ab}{}_{0i}\, ,\vphantom{\ds\int} & \ \quad \  & \Phi(\gamma){^A{}_{ij}} & = & P(\gamma){^A{}_{0ij}}\, ,\vphantom{\ds\int}\\
\Phi(\omega){}^{ab} & = &P(\omega){^{ab}{}_0}\, ,\vphantom{\ds\int} & & \Phi(F)^{i} & = & \cS(F){^{i }}-\partial_j P(B)^{ij}\, , \vphantom{\ds\int}\\
\Phi(B){{}_i} & = &P(B){}_{0i}\, ,\vphantom{\ds\int} & & \Phi(R)^{ab i} & = & \cS(R){^{ab i }} - \nabla_j P(B)^{ab\,ij}\, , \vphantom{\ds\int}\\
\Phi(A) & = &P(A){{}_0}\, ,\vphantom{\ds\int} & & \Phi(\cG)^a & = & \cS(\cG)^a + \nabla_i P(e)^{a\,i}- \frac{1}{4}  \,\beta_{b\,ij}\,P(B)^{ab\,ij}\, ,\vphantom{\ds\int}\\
\Phi(e)^a & = &  P(e){^a{}_0}\, ,\vphantom{\ds\int} & & \Phi(\nabla e)^{a\,i} & = &  \cS(\nabla e){^{a\,i}} - \nabla_j P(\beta){}^{a\,ij} +  \frac{1}{2}\,e_{b\,j}\,P(B)^{ab\, ij}\, ,\vphantom{\ds\int}\\
\Phi(\beta){^a{}_i} & = &P(\beta){^a{}_{0i}}\, ,\vphantom{\ds\int} & & \Phi(\nabla \phi)^{A\,ij} & = & \cS(\nabla \phi)^{A\,ij}+\nabla_k P(\gamma){}^{A\,ijk}-\rhd_B{}^A\,\,\phi^B\,P(B){}^{ij}\, ,\vphantom{\ds\int}\\
\Phi(\nabla B)& = & \multicolumn{5}{l}{\ds \cS(\nabla B) + \partial_i P(A)^i + \frac{1}{3!}\,\gamma^{A}{}_{ijk}\,\rhd_A{}^B\,P(\gamma){}_{B}{}^{ijk}-\phi_A\,\rhd_B{}^A\,P(\phi)^B\, ,\vphantom{\ds\int} } \\
\Phi(\nabla B)^{ab}& = & \multicolumn{5}{l}{ \cS(\nabla B)^{ab} + \nabla_i P(\omega)^{ab}{}^i +B{}^{[a|}{}_{c\,ij}\, P(B)^{c|b]\,ij}- 2e{^{[a|}{}_i}\, P(e)^{|b]\,i}- \beta{^{[a|}{}_{ij}}\, P(\beta)^{|b]\,ij}\, ,\vphantom{\ds\int} } \\
\end{array}
\end{equation}
are the first-class constraints, while
\begin{equation}
\begin{array}{lclclcl}
\chi(B){_{ab}{}^{jk}}=P(B){_{ab}{}^{jk}}\, , & & \chi(B){{}^{jk}}=P(B){{}^{jk}}\, , & & \chi(e){_a{}^i}=P(e){_a{}^i}\, , & & \chi(\phi)_A=P(\phi)_A\, , \vphantom{\ds\int} \\ 
\chi(\omega){_{ab}{}^i}=P(\omega){_{ab}{}^i}\, , & & \chi(A){{}^i}=P(A){{}^i}\, , & & \chi(\beta){_a{}^{ij}}=P(\beta){_a{}^{ij}}\, , & & \chi(\gamma){_A{}^{ijk}}=P(\gamma){_A{}^{ijk}}\, , \vphantom{\ds\int} \\
\end{array}
\end{equation}
are the second-class constraints. 

The PB algebra of the first-class constraints is given by:
\begin{equation} \label{FirstClassPBalgebra}
\begin{array}{lcl}
\pb{\Phi(\cG)^a (x)}{\Phi(\nabla e)_{b}{}^{ i}(y)} & = & - \,\Phi(R)^a{}_b{}^{i}(x)\, \delta^{(3)}(\vec{x}-\vec{y}) \, ,\vphantom{\ds\int}  \\
\pb{\Phi(\cG)^a (x)}{\Phi(\nabla B)_{bc}(y)} & = &  2\delta{}^a{}_{[b|} \, \Phi(\cG)_{|c]}(x)\, \delta^{(3)}(\vec{x}-\vec{y}) \, , \vphantom{\ds\int} \\
\pb{\Phi(\nabla e){^a{}_i}(x)}{\Phi(\nabla B)_{bc}(y)} & = & 2 \delta{}^a{}_{[b|} \Phi(\nabla e){_{|c]i}}(x)\,\delta^{(3)}(\vec{x}-\vec{y})\, ,\vphantom{\ds\int}\\
\pb{\Phi(R){^{abi}}(x)}{\Phi(\nabla B)_{cd}(y)} & = &-4\delta{}^{[a|}{}_{[c}\,\Phi(R){^{|b]}{}_{d]}{}^i}(x)\,\delta^{(3)}(\vec{x}-\vec{y})\, , \vphantom{\ds\int}\\
\pb{\Phi(\nabla B)^{ab}(x)}{\Phi(\nabla B)_{cd}(y)} & = &-4 \delta{}^{[a|}{}_{[c|} \, \Phi(\nabla B)^{|b]}{}_{|d]} (x)\,\delta^{(3)}(\vec{x}-\vec{y})\, ,\vphantom{\ds\int}\\
\pb{\Phi(\nabla B)(x)}{\Phi(\nabla \phi)_A{}^{ij}(y)}&=&-2 \rhd^B{}_A\,\,\Phi(\nabla \phi){}_B{}^{ij}(x)\delta^{(3)}(\vec{x}-\vec{y})\, .\vphantom{\ds\int} \\
\end{array}
\end{equation}
The PB algebra between the first and the second-class constraints is given by:
 \begin{equation}
\begin{array}{lcl}
\pb{\Phi(R)^{abi}(x)}{\chi(\omega)_{cd}{}^j(y)} & = & 4\,\delta{}^{[a|}{}_{[c|} \,\chi(B)^{|b]}{}_{|d]}{}^{ij}(x) \delta^{(3)}(\vec{x}-\vec{y}) \, , \vphantom{\ds\int} \\
\pb{\Phi(\cG)^a(x)}{\chi(\omega)_{cd}{}^i(y)} & = & 2\, \delta{}^a{}_{[c|} \,\chi(e)_{|d]}{}^i(x) \delta^{(3)}(\vec{x}-\vec{y}) \, , \vphantom{\ds\int} \\
\pb{\Phi(\cG)^a(x)}{\chi(\beta)_c{}^{jk}(y)} & = & \ds - \frac{1}{2}\, \chi(B)^{a}{}_{c}{}^{jk}(x)\, \delta^{(3)}(\vec{x}-\vec{y}) \, , \vphantom{\ds\int} \\
\pb{\Phi(\nabla e)^{ai}(x)}{\chi(\omega)_{cd}{}^j(y)} & = & - 2\,\delta{}^a{}_{[c|}\, \chi(\beta)_{|d]}{}^{ij}(x)\, \delta^{(3)}(\vec{x}-\vec{y}) \, , \vphantom{\ds\int} \\
\pb{\Phi(\nabla e)^{ai}(x)}{\chi(e)_b{}^j(y)} & = & \ds \frac{1}{2}\, \chi(B)^a{}_b{}^{ij} \,\delta^{(3)}(\vec{x}-\vec{y}) \, , \vphantom{\ds\int} \\
\pb{\Phi(\nabla B)^{ab}(x)}{\chi(\omega)_{cd}{}^i(y)} & = & 4\,\delta{}^{[a|}{}_{[c|}\, \chi(\omega)_{|d]}{}^{|b]}{}^i\, \delta^{(3)}(\vec{x}-\vec{y}) \, , \vphantom{\ds\int} \\
\pb{\Phi(\nabla B)(x)}{\chi(A){}^i(y)} & = & 2\, \chi(A){}^i\, \delta^{(3)}(\vec{x}-\vec{y}) \, , \vphantom{\ds\int} \\
\pb{\Phi(\nabla B)^{ab}(x)}{\chi(\beta)_c{}^{jk}(y)} & = & -2 \delta{}^{[a|}{}_c\, \chi(\beta)^{|b]jk}\, \delta^{(3)} (x-y)\, , \vphantom{\ds\int} \\
\pb{\Phi(\nabla B)(x)}{\chi(\gamma)_A{}^{ijk}(y)} & = & \rhd_A{}^B\,\, \chi(\gamma)_{B}{}^{ijk}(x)\, \delta^{(3)}(\vec{x}-\vec{y}) \, , \vphantom{\ds\int} \\
\pb{\Phi(\nabla B)^{ab}(x)}{\chi(B)_{cd}{}^{jk}(y)} & = & 4\,\delta{}^{[a|}{}_{[c}\, \chi(B)_{d]}{}^{|b]jk}\, \delta^{(3)}(\vec{x}-\vec{y}) \, , \vphantom{\ds\int} \\
\pb{\Phi(\nabla B)^{ab}(x)}{\chi(e)_{a}{}^{i}(y)} & = & -2 \delta{}^{[a|}{}_c \,\chi(e)^{|b]i}\, \delta^{(3)}(\vec{x}-\vec{y}) \, , \vphantom{\ds\int} \\
\pb{\Phi(\nabla B)(x)}{\chi(\phi)_{A}{}(y)} & = & -\rhd^B{}_A\, \chi(\phi){}_{B}(x)\, \delta^{(3)}(\vec{x}-\vec{y}) \, , \vphantom{\ds\int} \\
\pb{\Phi(\nabla \phi){^A{}^{ij}}(x)}{\chi(A){}^{k}(y)}&=&-\rhd_B{}^A \,\chi(\gamma)^B{}^{ijk}(x)\, \delta^{(3)}(\vec{x}-\vec{y})\, ,\vphantom{\ds\int}\\
\pb{\Phi(\nabla \phi){^A{}^{ij}}(x)}{\chi(\phi)_B{}(y)}&=&-\rhd_B{}^A\, \chi(B){}^{ij}(x)\, \delta^{(3)}(\vec{x}-\vec{y})\, .\vphantom{\ds\int}\\
\end{array}
\end{equation}
 The PB algebra between the second-class constraints has already been calculated, and is given in Equations~(\ref{AlgebraPrimarnihVeza}).

\section{The Bianchi Identities\label{section3}}
In order to calculate the number of degrees of freedom in the theory, one needs to make use of the \emph{Bianchi identities} (BI), as well as additional, \emph{generalized Bianchi identities} (GBI) that are an analogue of the ordinary BI for the additional fields present in the theory.

One uses BI associated with the $1$-form fields $\omega^{ab}$ and $e^a$, as well as the GBI for the $1$-form $A$. Namely, the corresponding $2$-form curvatures
\begin{equation}
\begin{array}{c}
R^{ab} = \D\omega^{ab} + \omega^a{}_c\wedge\omega^{cb} \, , \qquad T^a= \D e^a +\omega^a{}_b\wedge e^b\, , \qquad F=\D A\, , \vphantom{\ds\int} 
\end{array}
\end{equation}
satisfy the following identities:
\begin{equation} 
\epsilon^{\lambda\mu\nu\rho} \,\nabla_\mu R^{ab}{}_{\nu\rho} = 0 \, ,\vphantom{\ds\int}\label{bia1}
\end{equation}
\begin{equation}\label{bic}
\epsilon^{\lambda\mu\nu\rho}\left( \nabla_\mu T^a{}_{\nu\rho} -\, R^{ab}{}_{\mu\nu}\,e_{b\rho }\right) = 0 \vphantom{\ds\int}\, ,
\end{equation}
\begin{equation} 
\epsilon^{\lambda\mu\nu\rho} \,\nabla_\mu F{}_{\nu\rho} = 0 \, .\vphantom{\ds\int}\label{bia2}
\end{equation}
Choosing the free index to be time coordinate $\lambda=0$, these indentities, as the time-independent parts of the Bianchi identities, become the off-shell restrictions in the sense of the Hamiltonian analysis. On the other hand, choosing the free index to be a spatial coordinate, one obtains time-dependent pieces of the Bianchi identities, which do not enforce any restrictions, but can instead be derived as a consequence of the Hamiltonian equations of motion.

There are also GBI associated with the $2$-form fields $B^{ab}$, $B$ and $\beta^a$. The corresponding $3$-form curvatures are given by
\begin{equation}
 S^{ab}= \D B^{ab} + 2\omega^{[a|}{}_c \wedge B^{c\,|b]} \, , \qquad  P= \D B \, , \qquad G^a = \D \beta^a + \omega^a{}_b \wedge \beta^b \, .\vphantom{\ds\int}
\end{equation}
Differentiating these expressions, one obtains the following GBI:
\begin{equation}
 \epsilon^{\lambda\mu\nu\rho}\left(\frac{1}{3} \nabla_\lambda \,S^{ab}{}_{\mu\nu\rho} - R^{[a|\,c}{}_{\lambda\mu}\,B_c{}^{|b]}{}_{\nu\rho}\right) = 0\, ,\vphantom{\ds\int} \label{b21c}
 \end{equation}
\begin{equation}
 \epsilon^{\lambda\mu\nu\rho}\partial_\lambda \,P{}_{\mu\nu\rho} = 0\, ,\vphantom{\ds\int} \label{b22c}
 \end{equation}
\begin{equation} 
\epsilon^{\lambda\mu\nu\rho}\left( \frac{2}{3}\nabla_\lambda \,G^a{}_{\mu\nu\rho} - \,R^{ab}{}_{\lambda\mu}\,\beta{}_{b\,\nu\rho}\right) = 0 \, .\vphantom{\ds\int}\label{b2d}
\end{equation}
However, in four-dimensional spacetime, these identities will be single-component equations, with no free spacetime indices, and therefore necessarily feature time derivatives of the fields. Thus, they do not impose any off-shell restictions on the canonical variables. 

Finally, there is also GBI associated with the $0$-form $\phi$. The corresponding $1$-form curvature is:
\begin{equation}
    Q^A = \D \phi^A + \rhd_B{}^A \, A \wedge \phi^B\, ,
\end{equation}
so that the GBI associated with this curvature is:
\begin{equation}\label{bid}
    \epsilon^{\lambda\mu\nu\rho}\left( \nabla_\nu Q^A{}_\rho - \frac{1}{2}\,\rhd_B{}^A\, F{}_{\nu\rho} \phi^B \right) =0.
\end{equation}
This GBI consists of $12$ component equations, corresponding to six possible choices of the free antisymmetrized spacetime indices $\lambda\mu$, and the $2$ possible choices of the free group index $A$. However, not all of these $12$ identities are independent. This can be seen by taking the derivative of the Equation~(\ref{bid}) and obtaining eight identities of the form
\begin{equation} \label{bidzvezda}
    \rhd_B{}^A\,\epsilon^{\lambda \mu \nu \rho}\,\partial_\mu\,F{}_{\nu\rho}\, \phi^B=0\, ,
\end{equation}
which are automatically satisfied because of the GBI~(\ref{bia2}). One concludes there are only four independent identities~(\ref{bid}). Now, fixing the value $\lambda=0$, one obtains the time-independent components of both Equations~(\ref{bid}) and~(\ref{bidzvezda}),
\begin{equation}\label{bidprostorni}
    \epsilon^{0ijk}\left( \nabla_j Q^A{}_k - \frac{1}{2}\,\rhd_B{}^A\, F{}_{jk} \phi^B \right) =0\, ,
\end{equation}
and
\begin{equation} \label{bidzvezdaprostorni}
    \rhd_B{}^A\,\epsilon^{0ijk}\,\partial_i\,F{}_{jk}\, \phi^B=0\, .
\end{equation}
Of these, there are six components in Equation~(\ref{bidprostorni}), but, because of the two components of Equation~(\ref{bidzvezdaprostorni}), there are overall only four independent GBI relevant for the Hamiltonian analysis.

\section{Number of Degrees of Freedom\label{section4}}

Let us now show that the structure of the constraints implies that there are no local degrees of freedom (DoF) in a $3BF$ theory. In the general case, if there are $N$ initial fields in the theory and there are  $F$ independent first-class constraints per space point and $S$ independent second-class constraints per space point, then the number of local DoF, i.e., the number of independent field components, is given by
\begin{equation} \label{ndof}
n = N - F - \frac{S}{2}\, .
\end{equation}
Equation~(\ref{ndof}) is a consequence of the fact that $S$ second-class constraints are equivalent to vanishing of $S/2$ canonical coordinates and $S/2$ of their momenta. The $F$ first-class constraints are equivalent to vanishing of $F$ canonical coordinates, and since the first-class constraints generate the gauge symmetries, we can impose $F$ gauge-fixing conditions for the corresponding $F$ canonical momenta. Consequently, there are $2N - 2F - S$ independent canonical coordinates and momenta and therefore $2n = 2N-2F-S$, giving rise to Equation~(\ref{ndof}).

In our case, $N$ can be determined from the table \ref{TableOne},
\begin{table}[!ht]
\begin{center}
\caption{\label{TableOne}The number of components for all fields present in the theory.}
  \begin{tabular}{cccccccc}\toprule
$\vphantom{\ds\int}\omega^{ab}{}_{\mu}$ & $A{}_{\mu}$ & $\beta^a{}_{\mu\nu}$ & $\gamma^A{}_{\mu\nu\rho}$ & $B^{ab}{}_{\mu\nu}$ & $B{}_{\mu\nu}$ & $e^a{}_{\mu}$ & $\phi^A$ \\ \midrule
$24$ & $4$ & $24$ & $8$ & $36$ & $6$ & $16$ & $2$ \\ \bottomrule
    \end{tabular}
\end{center}
\end{table}
giving rise to a total of $N=120$ canonical coordinates.  Similarly, the number of independent components for the second class constraints is determined by the table \ref{TableTwo},
\begin{table}[!ht]
\begin{center}
\caption{\label{TableTwo}The number of components for the second class constraints present in the theory.}
\begin{tabular}{cccccccc}\toprule
$\vphantom{\ds\int}\chi(B)_{ab}{}^{jk}$ & $\chi(B){}^{jk}$ & $\chi(e)_a{}^i$ &  $\chi(\phi)_A$ & $\chi(\omega)_{ab}{}^i$ & $\chi(A){}^i$ & $\chi(\beta)_a{}^{ij}$ & $\chi(\gamma)_A{}^{ijk}$ \\  \midrule
$18$ & $3$ & $12$ & $2$ & $18$ & $3$ & $12$ & $2$\\  \bottomrule
\end{tabular}
\end{center}
\end{table}
so that $S=70$.

The first-class constraints are not all independent because of BI and GBI. To see that, take the derivative of $\Phi(R)^{abi}$ to obtain
\begin{equation}
\nabla_i \Phi(R)^{abi} = \lc^{0ijk} \nabla_i R^{ab}{}_{jk} + \frac{1}{2}R^{c[a|}{}_{ij} P(B)_c{}^{|b]ij}\, .
\end{equation}
The first term on the right-hand side is zero off-shell because $\epsilon ^{ijk}\, \nabla_i R^{ab}{}_{jk} =0$, which is a $\lambda=0$ component of the BI~(\ref{bia1}). The second term on the right-hand side is also zero off-shell, since it is a product of two constraints,
\begin{equation}
R^{c[a|} {}_{ij} \,P(B)_c{}^{|b]ij} \equiv \frac{1}{2}\epsilon_{0ijk}\cS(R)^{c[a|}{}^{k} \,P(B)_c{}^{|b]ij} = 0\, .
\end{equation}
Therefore, we have the off-shell identity
\begin{equation} \label{OffShellIdentitetZaR}
\nabla_i \Phi(R)^{abi} = 0\, ,
\end{equation}
which means that six components of $\Phi(R)^{abi}$ are not independent of the others. In an analogous fashion, taking the derivative of $\Phi(F)^{i}$, one obtains
\begin{equation}
\partial_i \Phi(F)^{i} = \lc^{0ijk} \,\partial_iF{}_{jk} + \frac{1}{2}\,F_{ij}\, P(B){}^{ij}\, .
\end{equation}
The first term on the right-hand side is zero off-shell because $\epsilon ^{ijk}\, \partial_i F{}_{jk} =0$, which is a $\lambda=0$ component of the GBI~(\ref{bia1}). The second term on the right-hand side is also zero off-shell, since it is a product of two constraints,
\begin{equation}
F{}_{ij}\, P(B)^{ij} \equiv \frac{1}{2}\epsilon_{0ijk}\,\cS(F){}^{k} \,P(B){}^{ij} = 0\, .
\end{equation}
Therefore, we have the off-shell identity
\begin{equation} \label{OffShellIdentitetZaF}
\partial_i \Phi(F)^{i} = 0\, ,
\end{equation}
which means that one component of $\Phi(F)^{i}$ is not independent of the others. Similarly, one can demonstrate that
\begin{equation}\label{OffShellIdentitetZaT0}  
 \nabla_i \Phi(\nabla e)_a{}^{i} - \frac{1}{2}\, \Phi(R)_{ab}{}^i\, e^b{}_i +\frac{1}{4}\epsilon^{0ijk}\cS(R){}_{ab}{}_{k}\,P(\beta)^b{}_{ij}= \frac{1}{2}\epsilon^{0ijk}\left( \nabla_i T_{a jk} - R_{ab\,ij}\,e^b{}_k \right)\, . 
\end{equation}
The right-hand side of the Equation~(\ref{OffShellIdentitetZaT0}) is the $\lambda=0$ component of the BI~(\ref{bic}), so that Equation~(\ref{OffShellIdentitetZaT0}) gives the~relation:
\begin{equation}\label{OffShellIdentitetZaT}  
     \nabla_i \Phi(\nabla e)_a{}^{i} - \frac{1}{2}\, \Phi(R)_{ab}{}^i\, e^b{}_i =0\, ,
\end{equation}
where we have omitted the term that is the product of two constraints. This relation means that four components of the constraints $ \Phi(\nabla e)_a{}^{i}$ and $\Phi(R)_{ab}{}^i$ can be expressed in terms of the rest. Finally, one~can also demonstrate that
\begin{equation}\label{OffShellIdentitetZaD0}
\begin{aligned}
\nabla_i \Phi(\nabla \phi)_A{}^{ij}&- \frac{1}{2}\epsilon_{0ikl}\,\rhd^{}_A\,\cS(F){}^{l}\,\chi(\gamma)_B{}^{ijk} +   \rhd^B{}_A\, \phi_B\, \Phi(F){}^{j}\\&+\frac{1}{2}\epsilon_{0ilm}\rhd^B{}_A\,P(B){}^{ij}\, \cS(\nabla \phi)_B{}^{lm}=\epsilon^{0ijk}\left(\nabla_i Q_A{}_k+\frac{1}{2}\rhd^B{}_A\,F{}_{ik}\,\phi_B\right)\, ,
\end{aligned}
\end{equation}
which gives
\begin{equation}\label{OffShellIdentitetZaD}
\nabla_i \Phi(\nabla \phi)_A{}^{ij}+ \frac{1}{2} \rhd^B{}_A \,\phi_B\, \Phi(F){}^{j}=0\, ,
\end{equation}
for $\lambda=0$ component of the GBI~(\ref{bid}), where we have again used that the product of two contraints is zero off-shell. This relation suggests that six components of two first-class constraints, $\Phi(\nabla \phi)_A{}^{ij}$ and $\Phi(F){}^j$, are not independent of the others. However, in the previous section, we have discussed that only four of these six identities are mutually independent, which means that we have only four independent identities~(\ref{OffShellIdentitetZaD}). A rigorous proof of this statement entails the evaluation of the corresponding Wronskian, and is left for future work.

Taking into account all of the above indentites~(\ref{OffShellIdentitetZaR}), (\ref{OffShellIdentitetZaF}), (\ref{OffShellIdentitetZaT}), and (\ref{OffShellIdentitetZaD}), we can finally evaluate the total number of independent first-class constraints. From the table \ref{TableThree},
\begin{table}[!ht]
\begin{center}
\caption{\label{TableThree}The number of components for the first class constraints present in the theory. The identities~(\ref{OffShellIdentitetZaR}), (\ref{OffShellIdentitetZaF}), (\ref{OffShellIdentitetZaT}), and~(\ref{OffShellIdentitetZaD}) reduce the number of components which are independent. This reduction is explicitly denoted in the table.}
\setlength{\tabcolsep}{1pt}
\begin{tabular}{ccccccccccccccc} \toprule
$\vphantom{\ds\int}\Phi(B)_{ab}{}^i$ & $\Phi(B){}^i$ & $\Phi(e)_a$ & $\Phi(\omega)_{ab}$ & $\Phi(A)$ &$\Phi(\beta)_a{}^i$  & $\Phi(\gamma)_A{}^{ij}$ & $\Phi(R)_{ab}{}^{i}$ &  $\Phi(F){}^{i}$ & $\Phi(\cG)_a$ & $\Phi(\nabla e)_{a}{}^{i}$ & $\Phi(\nabla B)_{ab}$ & $\Phi(\nabla B)$ & $\Phi(\nabla \phi)_A{}^{ij}$ \\  \midrule
$18$ & $3$ & $4$ & $6$ & $1$ & $12$ & $6$ & $18-6$ & $3-1$ & $4$ & $12-4$ & $6$ & $1$ & $6-4$ \\ \bottomrule
\end{tabular}
\end{center}
\end{table}
one can see that the total number of components of the first-class constraints is given by $F^*=100$. However, the number of independent components of the first-class constraints is $F=85$, obtained by subtracting the six relations~(\ref{OffShellIdentitetZaR}), one relation~(\ref{OffShellIdentitetZaF}), four relations~(\ref{OffShellIdentitetZaT}) and four relations~(\ref{OffShellIdentitetZaD}).

Therefore, substituting all the obtained results into Equation~(\ref{ndof}), one gets
\begin{equation} \label{BrojFizickihStepeniSlobode}
n = 120-85-\frac{70}{2}= 0,
\end{equation}
which means that there are no propagating DoF in a $3BF$ theory described by the action~(\ref{eq:LagrangianAction}).

\section{Generator of the Gauge Symmetry\label{section5}}

Based on the results of the Hamiltonian analysis of the action~(\ref{eq:LagrangianAction}), it can also be interesting to calculate the generator of the complete gauge symmetry of the action. The gauge generator of the theory is obtained by using the Castellani's procedure (see Chapter V in \cite{blagojevic2002gravitation} for details of the~procedure), and~one gets the following result (see Appendix \ref{ApendiksC} for details of the calculation):
\vspace{6pt} 
\begin{equation}\label{eq:generator}
\begin{array}{ccl}
    G&=&\ds \int_{\Sigma_3}\D^3 \vec{x} \bigg( \frac{1}{2}( \nabla_0 \epsilon^{ab}{}_i)\Phi(B){}_{ab}{}^i-\frac{1}{2}\epsilon^{ab}{}_i\Phi(R)_{ab}{}^i+\frac{1}{2}(\nabla_0 \epsilon^{ab})\Phi(\omega){}_{ab}-\frac{1}{2}\epsilon^{ab}\Phi(\nabla B){}_{ab}\\ \ds& &\qquad \vphantom{\ds\int}+
    (\partial_0 \epsilon{}_i)\Phi(B){}^i-\epsilon{}_i\Phi(F){}^i+(\partial_0 \epsilon)\Phi(A)-\epsilon\Phi(\nabla B)\\ \ds& &\qquad\vphantom{\ds\int}+(\nabla_0\epsilon{}^{a})\Phi(e)_a-\epsilon^a \Phi(\cG)_a +( \nabla_0 \epsilon^a{}_i)\Phi(\beta){}_a{}^i- \epsilon^a{}_i \Phi(\nabla e)_a{}^i\\ & &\qquad\vphantom{\ds\int}+ \ds \frac{1}{2}(\nabla_0\epsilon{}^A{}_{ij})\Phi(\gamma){}_A{}^{ij}- \frac{1}{2}\epsilon^A{}_{ij}\Phi(\nabla \phi){}_A{}^{ij}\\  & & \qquad \vphantom{\ds\int}\ds+\epsilon^{ab}\left( \beta_{[a|0i}P(\beta)_{|b]}{}^i+e_{[a|0}P(e)_{|b]}+B_{[a|c0i}P(B)^c{}_{|b]}{}^i\right)-\epsilon\,\gamma_{A0ij}\,\rhd_B{}^A \,P(\gamma)^{Bij}\\& & \qquad \vphantom{\ds\int}\ds+\epsilon^a\beta_{b0i}P(B)^{abi}+\epsilon^a{}_i\,e_{b0}P(B)_a{}^{bi}\bigg)\, .
\end{array}
\end{equation}
Here, $\epsilon^{ab}{}_i$, $\epsilon^{ab}$, $\epsilon_i$, $\epsilon$, $\epsilon^a$, $\epsilon^a{}_i$ and $\epsilon^A{}_{ij}$ are the independent parameters of the gauge transformations.

Furthermore, one can employ the gauge generator to calculate the form-variations for all canonical coordinates and their corresponding momenta, by computing the Poisson bracket of the chosen variable $A(t,\vec{x})$ and the generator~(\ref{eq:generator}):
\begin{equation}
\delta_0 A(t,\vec{x}) = \poisson{A(t,\vec{x})}{G}\, .
\end{equation}
The results are given as follows:
$$
\begin{array}{lclclcl}
    \delta_0 \omega^{ab}{}_0 & = & \nabla_0\epsilon^{ab}\, ,&\quad & \delta_0 \pi(\omega)_{ab}{}^0 &=&-2\epsilon_{[a|}{}^{c}{}_i\pi(B)_{c|b]}{}^{0i}-2\epsilon_{[a|}{}^{c}\pi(\omega)_{c|b]}{}^0\, ,\vphantom{\ds\int}\\
    & & & & & & +2\epsilon_{[a|}\pi(e)_{|b]}{}^0+2\epsilon_{[a|i}\pi(\beta)_{|b]}{}^{0i}\, ,\vphantom{\ds\int}\\
    \delta_0 \omega^{ab}{}_i & = & \nabla_i\epsilon^{ab}\, ,&\quad&\delta_0 \pi(\omega)_{ab}{}^{i} &=&-2\epsilon_{[a|}{}^{c}{}_j\,\pi(B)_{c|b]}{}^{ij}-2\epsilon_{[a|}{}^{c}{}_i\,\pi(\omega)_{|b]c}{}^i\vphantom{\ds\int}\\
    & & & & & & +2\epsilon_{[a|} \,\pi(e)_{|b]i}+2\epsilon_{[a|}{}_j\pi(\beta)_{|b]}{}^{ij}\vphantom{\ds\int}\\
    &&&&&& + 2\epsilon^{0ijk}\,\nabla_{[j|}\epsilon_{ab}{}_{|k]}+\epsilon^{0ijk}\epsilon_{[a|}\beta_{|b]}{}_{jk}\, ,\vphantom{\ds\int}\\
    \delta_0 B^{ab}{}_{0i} & = & \nabla_0 \epsilon^{ab}{}_i+\epsilon^{[a|}{}_ie^{|b]}{}_0\,&\quad&\delta_0\pi(B)_{ab}{}^{0i}&=&2\epsilon_{[a|c}\,\pi(B)_{|b]}{}^c{}^i\, ,\vphantom{\ds\int}\\
    & & +2\epsilon^{[a|c}B{}^{|b]}{}_c{}_{0i}+\epsilon^{[a|}\beta^{|b]}{}_{0i}\, ,&\quad& & &\vphantom{\ds\int}\\
    \delta_0 B^{ab}{}_{ij} & = &2\nabla_{[i|}\epsilon^{ab}{}_{|j]}+2\epsilon^{[a|c}B^{|b]}{}_{cij}&\quad&\delta_0\pi(B)_{ab}{}^{ij}&=&2\epsilon_{[a|c}\,\pi(B)_{|b]}{}^{cij}\, ,\vphantom{\ds\int}\\
    & &+2\epsilon^{[a|}{}_{[i}e^{|b]}{}_{j]}+\epsilon^{[a|}\beta^{|b]}{}_{ij}\, ,&&&&\vphantom{\ds\int}\\
    \delta_0 A{}_0 & = & \partial_0\epsilon\, ,&\quad & \delta_0 \pi(A){}^0 &=&-\frac{1}{2}\epsilon^A{}_{ij}\,\rhd^B{}_A\,\pi(\gamma)_B{}^{0ij}\, ,\vphantom{\ds\int}\\
    \delta_0 A{}_i & = & \partial_i\epsilon\, ,&\quad&\delta_0 \pi(A){}^{i} &=& \epsilon^{0ijk}\partial_j\epsilon_k-\frac{1}{2}\epsilon^A{}_{jk}\,\rhd^B{}_A\,\pi(\gamma)_B{}^{ijk} \, ,\vphantom{\ds\int}\\
    \delta_0 B{}_{0i} & = & \partial_0 \epsilon{}_i\, ,&\quad&\delta_0\pi(B){}^{0i}&=&0\, ,\vphantom{\ds\int}\\
    \delta_0 B{}_{ij} & = &2\,\partial_{[i|}\epsilon_{|j]}+\epsilon^A{}_{ij}\,\rhd^B{}_A\,\phi_B\, ,&\quad&\delta_0\pi(B){}^{ij}&=&-\epsilon^{0ijk}\partial_k\epsilon\, ,\vphantom{\ds\int}\\
    \delta_0 \beta^a{}_{0i} &=& \nabla_0\epsilon^a{}_i-\epsilon^{ab}\beta_{b0i}\, ,&\quad& \delta_0 \pi(\beta)_{a}{}^{0i}&=&-\epsilon_{ab}\pi(\beta)^b{}^{0i}+\frac{1}{2}\epsilon^b\pi(B)_{ab}{}^{0i}\, ,\vphantom{\ds\int}\\
    \delta_0 \beta^a{}_{ij} &=&2\nabla_{[i|}\epsilon^a{}_{|j]}-\epsilon^{ab}\,\beta_{bij}\, ,&\quad&\delta_0 \pi(\beta)_{a}{}^{ij}&=&-\epsilon_{ab}\,\pi(\beta)^b{}^{ij}+\frac{1}{2}\epsilon^b\,\pi(B)_{ab}{}^{ij}\vphantom{\ds\int}\\
    & & & & & &-\epsilon^{0ijk}\,\nabla_k\epsilon^a\, ,\vphantom{\ds\int}\\
    \delta_0 e^a{}_0 & =& \nabla_0\epsilon^a-\epsilon^{ab}\,e_{b0}\, ,&\quad&\delta_0 \pi(e)_{a}{}^0&=&-\epsilon_{ab}\,\pi(e)^{b0}+\frac{1}{2}\epsilon^b{}_i\,\pi(B)_{ab}{}^{0i}\, ,\vphantom{\ds\int}\\
    \delta_0 e^a{}_i &=&\nabla_i\epsilon^a-\epsilon^{ab}\,e_{bi}\, ,&\quad&\delta_0 \pi(e)_{a}{}^{i}&=&-\epsilon_{ab}\,\pi(e)^b{}^i+\epsilon^{0ijk}\Big(\,\nabla_{[j|}\epsilon_a{}_{|k]}+\epsilon_{ab}\beta^{bjk}\Big)\,\vphantom{\ds\int}\\
    & & & & & &+\frac{1}{2}\epsilon^b{}_j\,\pi(B)_{ab}{}^{ij}\, ,\vphantom{\ds\int}\\
\end{array}
$$
\begin{equation}
\begin{array}{lclclcl}
    \delta_0 \gamma^A{}_{0ij} &=&\nabla_0\epsilon^A{}_{ij}-\epsilon\,\gamma^{B}{}_{0ij}\,\rhd^A{}_B\, ,&\quad&\delta_0 \pi(\gamma)_A{}^{0ij}&=&\epsilon\, \rhd^B{}_A\,\pi(\gamma)_B{}^{0ij}\, ,\vphantom{\ds\int}\\
    \delta_0 \gamma^A{}_{ijk} &=&-\,\epsilon\,\gamma^B{}_{ijk}\,\rhd_B{}^A+\nabla_{i}\epsilon^A{}_{jk}&\quad&\delta_0 \pi(\gamma)_A{}^{ijk}&=&\epsilon\,\rhd_A{}^B\left(\pi(\gamma)_B{}^{ijk}+\epsilon^{0ijk}\,\phi_B\right)\, ,\vphantom{\ds\int}\\
    & & -\nabla_{j}\epsilon^A{}_{ik}+\nabla_{k}\epsilon^A{}_{ij}\, ,& & & &\vphantom{\ds\int}\\
    \delta_0 \phi^A &=& \epsilon\,\phi^B\,\rhd{}^A{}_B\, ,&\quad&\delta_0 \pi(\phi)_A&=& \ds -\epsilon\,\rhd^B{}_A\,\pi(\phi)_B+\frac{1}{3!}\,\epsilon\,\epsilon^{0ijk}\,\rhd^B{}_A\,\gamma_B{}_{ijk} \vphantom{\ds\int}\\
    & & & & & & \ds -\frac{1}{2}\rhd_A{}_B\,\epsilon^B{}_{ij}\,\pi(B)^{ij}-\frac{1}{2}\epsilon^{0ijk}\,\nabla_i\epsilon^A{}_{jk}\, ,\vphantom{\ds\int}\\
\end{array}
\end{equation}
These transformations are an extension of the form-variations in the case of the Poincar\' e $2$-group obtained in \cite{Oliveira:2018upx}.

\section{Conclusions\label{section6}}
Let us summarize the results of the paper. In Section \ref{section1}, we have demonstated in detail how to use the idea of a categorical ladder to introduce the $3$-group structure corresponding to the theory of scalar electrodynamics coupled to Einstein--Cartan gravity. We have introduced the topological $3BF$ action corresponding to this choice of a $3$-group, as well as the constrained $3BF$ action which gives rise to the standard equations of motion for the scalar electrodynamics. In order to perform the canonical quantization of this theory, the complete Hamiltonian analysis of the full theory with constraints has to be performed, but the important step towards this goal is the Hamiltonian analysis of the topological $3BF$ action. This has been done in Section \ref{section2}. Here, the first-class and second-class constraints of the~theory, as well as their Poisson brackets, have been obtained. In Section \ref{section3}, we have discussed the Bianchi identities and also the generalized Bianchi identities, since they enforce restrictions in the sense of Hamiltonian analysis, and reduce the number of independent first-class constraints present in the theory. With this background material in hand, in Section \ref{section4}, the counting of the dynamical degrees of freedom present in the theory has been performed and it was established that the considered $3BF$ action is a topological theory, i.e., the diffeomorphism invariant theory without any propagating degrees of freedom. In Section \ref{section5}, we have constructed the generator of the gauge symmetries for the~theory, and we found the form-variations for all the variables and their canonical momenta.

The results obtained in this paper represent the straightforward generalization of Hamiltonian analysis done in \cite{Mikovic2015} for the Poincar\' e $2$-group, and a first example of the Hamiltonian analysis of a $3BF$ action. The fact that the theory was found to be topological is nontrivial, since it relies on the existence of the generalized Bianchi identities, which have been identified for the first time. In~addition to that, it~was demonstrated that the algebra of constraint closes, which is an important consistency check for the theory. There is another very interesting aspect of the constraint algebra. Namely, one can recognize, looking at the structure of Equations~(\ref{FirstClassPBalgebra}) that the subalgebra generated by the first-class constraint $\Phi(\nabla\phi){}_A{}^{ij}$ is in fact an \emph{ideal} of the constraint algebra because the Poisson bracket between this constraint and all other constraints is again proportional to that constraint. It is curious that precisely the constraint $\Phi(\nabla\phi){}_A{}^{ij}$ is the only one related to the Lie group $L$ from the $3$-group, according to its index structure, and also that the structure constant of the ideal is determined by the action $\rhd$ of the group $G$ on $L$. Let~us also note that the action $\rhd$ appears as well in the structure constants of the algebra between the first-class and second-class constraints.

The results of this work open several avenues for future research. From the point of view of mathematics, the relationship between the algebraic structures mentioned above should be understood in more detail. More generally, one should understand the correspondence between the gauge group generated by the generator~(\ref{eq:generator}) and the $3$-group structure used to define the theory. This is not viable in the special case of the $3$-group discussed in this work, but instead needs to be done in the case of a generic $3$-group, where homomorphisms $\delta$ and $\partial$ and the Peiffer lifting $\poisson{\_}{\_}$ are nontrivial. From the point of view of physics, the obtained results represent the fundamental building blocks for the construction of the quantum theory of scalar electrodynamics coupled to gravity, as well as a convenient model to discuss before proceeding to the Hamiltonian analysis and canonical quantization of the full Standard Model coupled to gravity, formulated as a $3BF$ action with suitable constraints \cite{Radenkovic2019}. Both the Hamiltonian analysis of constrained $3BF$ models and the corresponding canonical quantization programme need to be further developed in order to achieve these goals. Our work is a first step in this direction.

Finally, let us note in the end that the above list of topics for future research is by no means complete, and there are potentially many other interesting topics that can be studied in this context.

\vspace{6pt} 

\authorcontributions{Investigation, T.R. and M.V; methodology, T.R. and M.V.; writing--original draft preparation, T.R.; writing--review and editing, M.V. All authors have read and agreed to the published version of the manuscript.}
\funding{This work was supported by the project ON171031 of the Ministry of Education, Science and Technological Development (MPNTR) of the Republic of Serbia, and partially by the bilateral scientific cooperation between Austria and Serbia through the project ``Causality in Quantum Mechanics and Quantum Gravity-2018-2019'', No. 451-03-02141/2017-09/02, supported by the Federal Ministry of Science, Research and Economy (BMWFW) of the Republic of Austria, and the Ministry of Education, Science and Technological Development (MPNTR) of the Republic of Serbia.}


\conflictsofinterest{The authors declare no conflict of interest.} 

\abbreviations{The following abbreviations are used in this manuscript:
\vspace{6pt} 
\\
\noindent 
\begin{tabular}{@{}ll}
LQG & Loop Quantum Gravity\\
BI & Bianchi Identities\\
GBI & Generalized Bianchi Identities \\
DoF & Degrees of Freedom\\
PB & Poisson Bracket
\end{tabular}}

\appendixtitles{yes} 
\appendix
\section{\label{ApendiksA} The Equations of Motion for the Scalar Electrodynamics}
The action of scalar electrodynamics coupled to Einstein--Cartan gravity is given in the form~(\ref{eq:scalarelectrodynamics}):
\begin{equation}\label{eq:scalarelectrodynamics1}
\begin{aligned}
 S =\int_{\cM_4} & B^{ab}\wedge R_{ab} + B \wedge F + e_a\wedge \nabla \beta^a + \phi_A \, \nabla \gamma^A \vphantom{\ds\int} \\
 &- \lambda_{ab} \wedge \Big(B^{ab}-\frac{1}{16\pi l_p^2}\varepsilon^{abcd} e_c \wedge e_d\Big)\vphantom{\ds\int} \\
 &+ {\lambda}^A\wedge \Big(\gamma_A - \frac{1}{2} H_{abcA} e^a \wedge e^b \wedge e^c\Big) +\Lambda^{abA}\wedge \Big( H_{abcA}\varepsilon^{cdef}e_d\wedge e_e \wedge e_f- \nabla \phi_A \wedge e_a \wedge e_b\Big) \vphantom{\ds\int} \\
 &+ \lambda\wedge \Big(B-\frac{12}{q}{M_{ab}}e^a\wedge e^b\Big) + {\zeta^{ab}}\Big( {M_{ab}}\varepsilon_{cdef}e^c\wedge e^d \wedge e^e \wedge e^f- F \wedge e_a \wedge e_b \Big)\vphantom{\ds\int} 
 \\
 &-\frac{1}{2\cdot 4!} m^2\phi_A\,\phi^A \varepsilon_{abcd}e^a\wedge e^b \wedge e^c \wedge e^d\vphantom{\ds\int}\, .
\end{aligned}
\end{equation}

Varying the total action~(\ref{eq:scalarelectrodynamics}) with respect to the variables $B_{ab}$, $B$, $\omega_{ab}$, $\beta_a$, $\lambda_{ab}$, $\Lambda^{abA}$, $\gamma^A$, ${\lambda^A}$, $H_{abcA}$, ${\zeta^{ab}}$, ${M_{ab}}$,  $\lambda$, $A$, $\phi^A$ and $e^a$, one obtains the equations of motion:
\begin{equation}
\label{eq:1}
R^{ab}-\lambda^{ab}=0\, ,\vphantom{\ds\int}
\end{equation}
\begin{equation}
\label{eq:2}F+\lambda=0\, , \vphantom{\ds\int}
\end{equation}
\begin{equation}
\label{eq:3}
\nabla B^{ab} - e^{[a|} \wedge \beta^{|b]} = 0\, ,\vphantom{\ds\int}
\end{equation}
\begin{equation}
\label{eq:4}
\nabla e^a = 0\, ,\vphantom{\ds\int}
\end{equation}
\begin{equation}
\label{eq:5} 
B^{ab}-\frac{1}{16\pi l_p^2}\varepsilon^{abcd} e_c \wedge e_d=0\, ,\vphantom{\ds\int}
\end{equation}
\begin{equation}
\label{eq:6} 
H_{abcA}\varepsilon^{cdef}e_d\wedge e_e \wedge e_f- \nabla \phi_A \wedge e_a \wedge e_b=0\, ,\vphantom{\ds\int}
\end{equation}
\begin{equation}
\label{eq:7} 
\nabla \phi_A-{\lambda}_A=0\, ,\vphantom{\ds\int}
\end{equation}
\begin{equation}
\label{eq:8} 
\gamma_A - \frac{1}{2} H_{abcA} e^a \wedge e^b \wedge e^c=0\, ,\vphantom{\ds\int}
\end{equation}
\begin{equation}
\label{eq:9} 
-\frac{1}{2}\lambda{}^A\wedge e^a \wedge e^b \wedge e^c+\varepsilon^{cdef}\Lambda^{abA}\wedge e_d\wedge e_e \wedge e_f=0\, ,\vphantom{\ds\int}
\end{equation}
\begin{equation}
\label{eq:10}{M_{ab}}\varepsilon_{cdef}e^c \wedge e^d \wedge e^e \wedge e^f- F \wedge e_a \wedge e_b=0\, , \vphantom{\ds\int}
\end{equation}
\begin{equation}
\label{eq:11}-\frac{12}{q}\lambda\wedge e^a \wedge e^b + \zeta^{ab} \varepsilon_{cdef}e^c \wedge e^d \wedge e^e \wedge e^f=0\, , \vphantom{\ds\int}
\end{equation}
\begin{equation}
\label{eq:12}B-\frac{12}{g}{M_{ab}}e^a\wedge e^b=0\, , \vphantom{\ds\int}
\end{equation}
\begin{equation}
\label{eq:13}
-\D B+\D(\zeta^{ab} e_a \wedge e_b)-\phi_A \rhd_B{}^A \gamma^B-\Lambda^{abA}\,\rhd^B{}_A\,\phi_B\wedge e_a\wedge e_b=0\, , \vphantom{\ds\int}
\end{equation}
\begin{equation}
\label{eq:14} 
\nabla \gamma_A-\nabla (\Lambda^{ab}{}_A \wedge e_a \wedge e_b)-\frac{1}{4!}  m^2 \,\phi_A \varepsilon_{abcd}e^a\wedge e^b \wedge e^c \wedge e^d=0\, ,\vphantom{\ds\int}
\end{equation}
\begin{equation}
\label{eq:15}
\begin{aligned}
    \nabla \beta_a&+\frac{1}{8\pi l_p^2}\varepsilon_{abcd}\lambda^{bc}\wedge e^d+\frac{3}{2} H_{abcA}\lambda^A\wedge e^b\wedge e^c+3H^{defA}\varepsilon_{abcd}\Lambda_{efA}\wedge e^b\wedge e^c\vphantom{\ds\int}\\&-2\Lambda_{abA}\wedge \nabla \phi^A \wedge e^b-2\frac{1}{4!}  m^2 \phi_A\, \,\phi^A \varepsilon_{abcd} e^b \wedge e^c \wedge e^d\\&- \frac{24}{q}{M_{ab}}\lambda\wedge e^b
+ 4{\zeta^{ef}}{M_{ef}}\varepsilon_{abcd}e^b\wedge e^c \wedge e^d - 2 {\zeta_{ab}} F\wedge e^b=0\, .\vphantom{\ds\int}
\end{aligned}
\end{equation}
The dynamical degrees of freedom are the tetrad fields $e^a$, the scalar field $\phi^A$, and the electromagnetic potential $A$, while the remaining variables are algebraically determined in terms of them. Specifically, Equations~(\ref{eq:1})--(\ref{eq:12}) give
\begin{equation}
\begin{array}{c}\label{eq:sys1}
\ds    \lambda_{ab}{}_{\mu\nu}=R_{ab}{}_{\mu\nu}\, , \qquad    \omega^{ab}{}_\mu=\triangle^{ab}{}_\mu\, , \qquad 
    \gamma^A{}_{\mu\nu\rho}=-\frac{1}{2e}\varepsilon_{\mu\nu\rho\sigma}\,\nabla^\sigma\phi^A\, , \vphantom{\ds\int} \\
\ds    \Lambda^{abA}{}_{\mu}=\frac{1}{12e}g_{\mu\lambda}\varepsilon^{\lambda\nu\rho\sigma}\nabla_\nu\phi^A\, e{}^a{}_{\rho} e{}^b{}_\sigma\, ,\vphantom{\ds\int} \qquad \beta^a{}_{\mu\nu}=0\, , \qquad B_{ab}{}_{\mu\nu}=\frac{1}{8\pi l_p^2}\varepsilon_{abcd}e^c{}_\mu e^d{}_\nu\, , \\
\ds H^{abcA}=\frac{1}{6e} \varepsilon^{\mu\nu\rho\sigma}\,\nabla_\mu\phi^A\, e^a{}_\nu e^b{}_\rho e^c{}_\sigma\, , \qquad \lambda^A{}_{\mu}=\nabla_{\mu}\phi^A\, ,\vphantom{\ds\int}\\ \ds
    \lambda_{\mu\nu}=F_{\mu\nu}\, , \quad
    B_{\mu\nu}=-\frac{1}{2eq}{\varepsilon_{\mu\nu\rho\sigma}} {F^{\rho \sigma}}\, ,\vphantom{\ds\int}\\ \ds
     M^{ab}=-\frac{1}{4e}\varepsilon^{\mu \nu \rho \sigma}F_{\mu \nu}\,e^a{}_\rho e^b{}_\sigma\, , \quad  {\zeta^{ab}}=\frac{1}{4eq}\varepsilon^{\mu \nu \rho \sigma}F_{\mu \nu}\,e^a{}_\rho e^b{}_\sigma\, .\vphantom{\ds\int}
\end{array}
\end{equation}
Note that from the Equations~(\ref{eq:3})--(\ref{eq:5}) it follows that $\beta^a=0$, as in the pure gravity case. The equation of motion~(\ref{eq:14}) reduces to the covariant Klein--Gordon equation for the scalar field coupled to the electromagnetic potential $A$,
\begin{equation}
\left(\nabla_\mu\nabla^\mu -m^2\right)\phi_A=0\, .
\end{equation}
From Equation~(\ref{eq:13}), we obtain the differential equation of motion for the field $A$:
\begin{equation}
     \nabla_\mu F^{\mu\nu}=j^\nu\, , \quad \quad j^\mu\equiv \frac{1}{2}\Big(\nabla^\nu \phi^A \rhd^B{}_A \phi_B - \phi_A \rhd_B{}^A \nabla^\nu \phi^B\Big)=iq\,\Big(\nabla \phi^*\,\phi-\phi^* \nabla \phi\Big)\, .
\end{equation}
Finally, the equation of motion~(\ref{eq:15}) for $e^a$ becomes:
\begin{equation}\label{eq:scalareomfore}
\begin{array}{c}
     \ds R^{\mu\nu}-\frac{1}{2}g^{\mu\nu} R=8\pi l_p^2 \; T^{\mu\nu}\, ,\vphantom{\ds\int} \\
     \ds T^{\mu\nu}\equiv\nabla^\mu \phi_A \, \nabla^\nu \phi^A -\frac{1}{2}g^{\mu\nu} \left(\nabla_\rho \phi_A\, \nabla^\rho \phi^A+m^2\phi_A\,\phi^A \right)-\frac{1}{4q}\left(F_{\rho \sigma}F^{\rho \sigma}g^{\mu \nu}+4F^{\mu \rho}{F_\rho}^{\nu} \right)\, .\vphantom{\ds\int}
\end{array}
\end{equation}
The system of Equations~(\ref{eq:1})--(\ref{eq:15}) is equivalent to the system of Equations~(\ref{eq:sys1})--(\ref{eq:scalareomfore}). 

\section{\label{ApendiksC} The Calculation of the Gauge Generator}

The gauge generator of the theory is obtained by the standard Castellani procedure (see \cite{blagojevic2002gravitation} for an introduction). One starts from the generic form for the generator,
\begin{equation}
\begin{array}{ccl}
    G&=&\ds \int_{\Sigma_3}\partial^3 \vec{x} \Big( \frac{1}{2}( \partial_0 \epsilon^{ab}{}_i)G_1{}_{ab}{}^i+\frac{1}{2}\epsilon^{ab}{}_iG_0{}_{ab}{}^i+\frac{1}{2}(\partial_0 \epsilon^{ab})G_1{}_{ab}+\frac{1}{2}\epsilon^{ab}G_0{}_{ab}\\ & &\qquad\vphantom{\ds\int}+ \ds
    (\partial_0 \epsilon{}_i)G_1{}^i+\epsilon{}_i G_0{}^i+(\partial_0 \epsilon)G_1+\epsilon G_0\\ & &\qquad\vphantom{\ds\int}\ds +(\partial_0\epsilon{}^{a})G_1{}_a+\epsilon^a G_0{}_a+( \partial_0 \epsilon^a{}_i)G_1{}_a{}^i + \epsilon^a{}_i G_0{}_a{}^i\\&&\qquad\vphantom{\ds\int} \ds + \frac{1}{2}(\partial_0\epsilon{}^A{}_{ij})G_1{}_A{}^{ij} + \frac{1}{2}\epsilon^A{}_{ij}G_0{}_A{}^{ij}\Big)\, ,
\end{array}
\end{equation}
where the generators $G_0$ and $G_1$ are obtained by the standard prescription \cite{blagojevic2002gravitation}:
\begin{equation}
\begin{array}{r}
     G_1=C_{PFC}\, ,\vphantom{\ds\int}\\ 
     G_0+\pb{G_1}{H_T}=C_{PFC}\, ,\vphantom{\ds\int}\\
     \pb{G_0}{H_T}=C_{PFC}\, ,\vphantom{\ds\int}
\end{array}
\end{equation}
where $C_{PFC}$ is a primary first-class constraint. 
For example, one choses $G_1{}_{ab}{}^i=\Phi(B){}_{ab}{}^i$. From the~conditions
\begin{equation}
\begin{array}{r}
    G_{0ab}{}^i+\pb{\Phi(B){}_{ab}{}^i}{H_T}=G_{0ab}{}^i+\Phi(R)_{ab}{}^i=C_{PFC}\, ,\vphantom{\ds\int}\\
    \pb{G_{0ab}{}^i}{H_T}=C_{PFC}{}^*=\pb{C_{PFC}-\Phi(R)_{ab}{}^i}{H_T}\, ,\vphantom{\ds\int}
\end{array}
\end{equation}
we solve for $G_{0ab}{}^i$ by determining $C_{PFC}$ from the second equation. Evaluating one PB, one can reexpress the second equation in the form:
\begin{equation}
\pb{C_{PFC}}{H_T}= C_{PFC}{}^* +2\omega_{[a|}{}^d{}_0\Phi(R)_{|b]d}{}^i=\pb{2\omega_{[a|}{}^d{}_0 P(B)_{|b]d}{}^i}{H_T}\, .
\end{equation}
From the second equality, we recognize that
\begin{equation}
C_{PFC} = 2\omega_{[a|}{}^d{}_0 P(B)_{|b]d}{}^i\, ,
\end{equation}
which can then be substituted into the first condition above, giving
\begin{equation}
    G_{0ab}{}^i=2\omega_{[a|}{}^d{}_0 \Phi(B)_{|b]d}{}^i-\Phi(R)_{ab}{}^i\, .
\end{equation}
One thus obtains $$\frac{1}{2}( \partial_0 \epsilon^{ab}{}_i)(G_1)_{ab}{}^i+\frac{1}{2}\epsilon^{ab}{}_iG_0{}_{ab}{}^i=\frac{1}{2}\nabla_0\epsilon^{ab}{}_i\Phi(B)_{ab}{}^i-\frac{1}{2}\epsilon^{ab}{}_i\Phi(R)_{ab}{}^i\, .$$
The other $G_0$ and $G_1$ terms are obtained in a similar way, and the generator~(\ref{eq:generator}) is derived.

\section{\label{ApendiksCnovi} Introduction to 3-Groups}

The notion of a {\em $3$-group} is usually introduced in the framework of higher category theory~\cite{BaezHuerta2011}. In category theory, every {\em group} can be understood as a {\em category} which has only one element, and~morphisms which are all invertible. The group elements are then individual morphisms that map the category element to itself, while the group operation is the categorical composition of the morphisms. In such a case, the axioms of the category guarantee the validity of all axioms of a group. This kind of construction can be generalized to {\em $2$-groups, $3$-groups} and, in general, {\em $n$-groups}. Namely, a $2$-group is by definition a $2$-category which has only one element, and whose morphisms and $2$-morhisms (i.e., morphisms between morphisms) are invertible. Similarly, a $3$-group is by definition a $3$-category which has only one element, while its morphisms, $2$-morphisms, and $3$-morphisms are~invertible.

The above definition of a $3$-group is very abstract, and while theoretically very important, in itself not very useful for practical calculations and applications in physics. Fortunately, there is a theorem of equivalence between $3$-groups and the so-called {\em $2$-crossed modules}, which are algebraic structures with more familiar properties \cite{Wang2014, martins2011}. For the applications in physics, attention focuses on the so-called {\em strict Lie $3$-groups}, and their corresponding differential (Lie algebra) structure, which corresponds to the {\em differential Lie $2$-crossed module}. Let us therefore give a brief overview of the latter.

A \emph{differential Lie $2$-crossed module } $(\mathfrak{l}\stackrel{\delta}{\to} \mathfrak{h} \stackrel{\partial}{\to}\mathfrak{g},\,\rhd,\,\{\_\; ,\,\_ \})$ is given by three Lie algebras $\mathfrak{g}$, $\mathfrak{h}$ and $\mathfrak{l}$, maps $\delta: \mathfrak{l} \to \mathfrak{h}$ and $\partial$: $\mathfrak{h} \to \mathfrak{g}$, together with a map called the Peiffer lifting,
\begin{equation}
\{ \_\;,\,\_\}:\mathfrak{h}\times \mathfrak{h} \to \mathfrak{l}\, ,
\end{equation}
and an action $\rhd$ of the algebra $\mathfrak{g}$ on all three algebras. 

Let us introduce the bases in the three algebras, $\tau_\alpha\in\mathfrak{g}$, $t_a\in\mathfrak{h}$ and $T_A\in \mathfrak{l}$, and structure constants in those bases, as follows:
\begin{equation}
    [\tau_\alpha\, ,\tau_\beta]=f_{\alpha\beta}{}^{\gamma}\tau_\gamma\, , \quad [t_a\, ,t_b]=f_{ab}{}^c t_c\, , \quad [T_A\,T_B]=f_{AB}{}^C T_C\, .
\end{equation}
Now, the maps $\partial$ and $\delta$ can be written as
\begin{equation}
    \partial(t_a)=\partial_a{}^\alpha\, \tau_\alpha\, ,\quad \quad \delta(T_A)=\delta_A{}^a\,t_a\, ,
\end{equation}
and the action of the algebra $\mathfrak{g}$ on $\mathfrak{g}$, $\mathfrak{h}$ and $\mathfrak{l}$ as:
\begin{equation}
    \tau_\alpha \rhd \tau_\beta = \rhd_{\alpha\beta}{}^\gamma\,\tau_\gamma\, ,\quad \tau_\alpha \rhd t_a = \rhd_{\alpha a}{}^b \, t_b\, ,\quad \tau_\alpha \rhd T_A=\rhd_{\alpha A}{}^B \, T_B\, .
\end{equation}
Finally, the Peiffer lifting can be encoded into coefficients ${X_{ab}}^A$ as:
\begin{equation}
    \{t_a, \, t_b\} = {X_{ab}}^A\, T_A\, .
\end{equation}

A differential Lie $2$-crossed module has the following properties (we write all equations in the abstract and their corresponding component forms, side by side):
\begin{enumerate}[leftmargin=10mm,labelsep=3mm]
    \item The action of the algebra $\mathfrak{g}$ on itself is via the adjoint representation, i.e., $\forall g,g_1\in \mathfrak{g}$:
\begin{equation}
    g\rhd g_1=[g,g_1]\, , \quad \quad \quad \rhd_{\alpha\beta}{}^\gamma= f_{\alpha \beta}{}^\gamma \, .
\end{equation}
\item The action of the algebra $\mathfrak{g}$ on algebras $\mathfrak{h}$ and $\mathfrak{l}$ is $\mathfrak{g}$-equivariant, i.e., $\forall g\in \mathfrak{g}$, $h\in\mathfrak{h}$, $l\in\mathfrak{l}$:
\begin{equation}
    \partial(g\rhd h)=g\rhd\partial(h)\, ,\quad\quad\quad \partial_a{}^\beta\, f_{\alpha \beta}{}^\gamma  = \rhd_{\alpha a}{}^b \,\partial_b{}^\gamma\, ,
\end{equation}
\begin{equation}
    \delta(g\rhd l)=g\rhd \delta(l)\, ,\quad \quad \quad \delta_A{}^a\,\rhd_{\alpha a}{}^b = \rhd_{\alpha A}{}^B\, \delta_B{}^b\, .
\end{equation}
\item The Peiffer lifting is a $\mathfrak{g}$-equivariant map, i.e., for every $g \in \mathfrak{g}$ and $h_1, h_2 \in \mathfrak{h}$:
\begin{equation}
    g\rhd\{h_1,h_2\}=\{g\rhd h_1,h_2\}+\{h_1,\,g\rhd h_2\}\, ,\quad \quad \quad {X_{ab}}^B \,{\rhd_{\alpha B}}^A={\rhd_{\alpha a}}^c\, {X_{cb}}^A+{\rhd_{\alpha b}}^c\, {X_{ac}}^A\, .
\end{equation}
\item For every $h_1,\,h_2 \in \mathfrak{h}$, the following identity holds:
\begin{equation}
    \delta(\left\{h_1,\,h_2  \right \})=\left [h_1\, ,h_2\right]-\partial(h_1)\rhd h_2\, ,\quad \quad \quad {X_{ab}}^A\, {\delta_A}^c=f_{ab}{}^c-{\partial_a}^\alpha\,\rhd_{\alpha b}{}^c \, .
\end{equation}
\item For all $l_1,\,l_2 \in \mathfrak{l}$, the following identity holds:
\begin{equation}
    [l_1,\,l_2]=\left\{\delta(l_1),\,\delta(l_2)\right \}\, ,\quad\quad\quad {f_{AB}}^C={\delta_A}^a\,{\delta_B}^b\, {X_{ab}}^C\, .
\end{equation}
\item For all $h_1,h_2,h_3\in\mathfrak{h}$:
\begin{equation}
    \begin{aligned}
    \left\{[h_1,\,h_2],\,h_3\right\}&=&\partial(h_1)\rhd\left\{h_2,\,h_3\right \}+\left\{h_1,\,[h_2,\,h_3]\right\}-\partial(h_2)\rhd\left\{h_1,\,h_3\right \}-\left\{h_2,\,[h_1,\,h_3]\right \}\, ,\\
    {f_{ab}}^d\,{X_{dc}}^B&=&{\partial_a}^\alpha\, {X_{bc}}^A\,{\rhd_{\alpha A}}^B+{X_{ad}}^B\,{f_{bc}}^d-{\partial_b}^\alpha\,{\rhd_{\alpha A}}^B\,{X_{ac}}^A-{X_{bd}}^B\,{f_{ac}}^d\, .
    \end{aligned}
\end{equation}
\item For all $h_1,h_2,h_3\in\mathfrak{h}$:
\begin{equation}
\begin{aligned}
    \left\{h_1,\,[h_2,\,h_3]\right \}&=&\left\{\delta\left\{h_1,\,h_2\right \},h_3\right \}-\left\{\delta\left\{h_1,\,h_3\right \},\,h_2\right \}\, ,\\
    {X_{ad}}^A\, {f_{bc}}^d&=&{X_{ab}}^B\,{\delta_B}^d\, {X_{dc}}^A-{X_{ac}}^B\,{\delta_B}^d {X_{db}}^A\, .
\end{aligned}
\end{equation}
\item For all $l\in\mathfrak{l}$ and $\forall h\in\mathfrak{h}$:
\begin{equation}
    \left\{\delta(l),\,h\right \}+\left\{h,\,\delta(l)\right \}=-\partial(h) \rhd l\, , \quad \quad \quad 2\,{\delta_A}^a\, {X_{\{ab\}}}^B=-{\partial_b}^\alpha\, {\rhd_{\alpha A}}^B\, .
\end{equation}
\end{enumerate}

Finally, when dealing with various algebra valued differential forms, one multiplies them as differential forms using the ordinary wedge product $\wedge$, and simultaneously as algebra elements using one of maps defined above. For example, the product with an action $\wedge^{\rhd}$ of the $\mathfrak{g}$-valued $n$-form $\rho$ on the $\mathfrak{h}$-valued $m$-form $\eta$ is defined as:
\begin{equation}
\begin{aligned}
    \rho \wedge^{\rhd} \eta &=&\frac{1}{n! m!}\, \rho^\alpha{}_{\mu_1\dots\mu_m}\,\eta^a{}_{\nu_1\dots\nu_n}\,\tau_\alpha\rhd t_a \,\D x^{\mu_1}\wedge\dots\D x^{\mu_m}\wedge \D x^{\nu_1}\wedge\dots\wedge \D x^{\nu_n}\\&=&\frac{1}{n! m!}\, \rho^\alpha{}_{\mu_1\dots\mu_m}\,\eta^a{}_{\nu_1\dots\nu_n}\,\rhd_{\alpha a}{}^b t_b \,\D x^{\mu_1}\wedge\dots\D x^{\mu_m}\wedge \D x^{\nu_1}\wedge\dots\wedge \D x^{\nu_n}\, .
\end{aligned}
\end{equation}

\reftitle{References}



\end{document}